\documentclass[preprint,12pt]{elsarticle}

\usepackage{amssymb}

\usepackage{algpseudocode}
\usepackage{algorithm}
\usepackage{amsmath}
\usepackage{subcaption}
\usepackage{multirow}
\usepackage{graphicx}

\journal{Nonlinear Analysis: Hybrid Systems}

\begin{document}

\begin{frontmatter}

\title{Distributed switched model-based predictive control for distributed large-scale systems with switched topology} 

\author[1]{Morteza Alinia Ahandani}\ead{alinia@tabrizu.ac.ir},    
\author[1]{Hamed Kharrati}\ead{kharrati@tabrizu.ac.ir},               
\author[1]{Farzad Hashemzadeh }\ead{hashemzadeh@tabrizu.ac.ir},  
\author[1]{Mahdi Baradarannia}\ead{mbaradaran@tabrizu.ac.ir},               

\address[1]{Department of Control Engineering, Faculty of Electrical and Computer Engineering, University of Tabriz, Tabriz, Iran}


\begin{abstract}                          
Distributed switched large-scale systems are composed by dynamically coupled subsystems, in which interactions among subsystems vary over time according to an exogenous input signal named switching signal. In this paper, we present a distributed robust switched model-based predictive control (DSwMPC) to control such systems that are subject to local state/input constraints. The proposed method guarantees stabilizing the origin of the whole closed-loop system and ensures the constraints satisfaction in the presence of an unknown switching signal. In the distributed model-based predictive control (DMPC) used in this work, by considering the interactions among subsystems as an additive disturbance, the effect of the switch is reflected on the dynamic equation, local, and consistency constraint sets of the nominal subsystems. In the DSwMPC, to compensate the effect of switching signal which creates a time-varying network topology, a robust tube-based switched model-based predictive control (RSwMPC) with switch–robust control invariant (switch-RCI) set as the target set robust to unknown mode switching is used as local controller. The scheme performance is assessed using three typical examples. In the first example, the switching times are unknown in prior, but the next neighborhood sets are assumed to be known in prior. In the second and third cases, both of them are supposed to be unknown in prior. The simulation results show that the input and state constraints are satisfied by the proposed DSwMPC at all times. They also validate that the closed-loop system converges to the origin. Also, to demonstrate the performance of the proposed DSwMPC more clearly, a comparison of the DSwMPC with a centralized SwMPC (CSwMPC) and a decentralized SwMPC (DeSwMPC) shows that the DSwMPC outperforms the DecSwMPC and also the shapes of response curves under the CSwMPC are very similar to those obtained by the DSwMPC.
\end{abstract}


\begin{highlights}
  \item The main contribution of this paper is to apply the Robust switched MPC controllers in a distributed fashion on the distributed large-scale systems with switched topology, input and state constraints, in which the subsystems interact with each other by states and inputs.

  \item The proposed method guarantees stabilizing the origin of the whole closed-loop system and ensures the constraints satisfaction in the presence of an unknown switching signal.

  \item By considering the interactions among subsystems as an additive disturbance in distributed switched large-scale systems, the effect of the switch in network topology is reflected on the dynamic equation, local, and consistency constraint sets of the nominal subsystems.

  \item In the proposed distributed robust switched model-based predictive control, the robust tube-based switched model-based predictive control with switch–robust control invariant set as the target set robust to unknown mode switching is used as local controller.

  \item The simulation results show that the input and state constraints are satisfied by the proposed DSwMPC at all times. They also validate that the closed-loop system converges to the origin.
\end{highlights}


\end{frontmatter}

\section{Introduction} \label{Sect1:Intro}
Distributed large-scale systems such as electric power systems, manufacturing systems, process plants, communication networks and swarms of robots are those of high-dimensional systems composed of interconnections of several lower-dimensional subsystems. In the distributed large-scale systems, the subsystems can interact with neighboring subsystems by both inputs, states, and outputs and they can have their local and/or global constraints and goals. A group of this kind of system is the systems with switched topology. Switched systems as a sub-class of hybrid dynamic systems can be used to model systems that are subject to known or unknown abrupt parameter variations such as synchronously switched linear systems, networks with periodically varying switchings, and sudden change of system structures due to various reasons. They appear naturally in the study of multi-rate sampled-data systems. Also, the switched systems can also be employed to describe the overall system of a single process controlled utilizing multi-controller switching such as the hybrid control scheme for nonholonomic systems, which are not stabilizable by means of any individual continuous state feedback controller \cite{Ahandani1}. Hence, switching among different system structures is an essential feature of many engineering and practical real-world systems \cite{Sun2}, especially in large-scale systems due to the existence of a wide variety of actuators, sensors and communication networks. In the distributed switched large-scale systems studied in this research, interactions among subsystems vary over time according to a switching signal. They also can be modeled as networked control systems with time-varying network topology.

A centralized control implementation for distributed large-scale systems is typically not realistic. In a decentralized control system, no information is exchanged between the controllers and every controller makes its own decision. Still in the distributed control systems, controllers are allowed to share their information with local controllers. This paper focuses on DMPC strategy. In a DMPC control structure, the overall system under control is divided into many interacted subsystems, where each of them is controlled using a separate local controller based on model-based predictive control (MPC) strategy. This class of controllers combines the advantage of a decentralized control structure, e.g., high flexibility and good error tolerance, and benefits of a centralized control structure, e.g., good global performance \cite{Li2016_6}. The online solving an optimization problem at each sampling time in MPC increases the computational burden. So, the standard centralized MPC cannot be implemented for plants with fast dynamics and/or a large number of control variables. A natural solution to this problem is to decompose the plant into smaller subsystems and then to design local controllers \cite{Hernandez8}. This can be considered as another advantage of the distributed implementation of MPC than its centralized version. Motivated by those mentioned above, the current research employs a distributed MPC (DMPC) for control of distributed switched large-scale systems.

The different versions of DMPC proposed in the literature \cite{Scattolini9}, \cite{Christofides10}, \cite{Maestre11} and \cite{Negenborn12} can be classified according to various criteria. The DMPC can be typically categorized based on the protocol of information exchange into non-iterative or iterative algorithms. In the DMPC with a non-iterative-based algorithm, each local MPC controller communicates only once per time-step with other local MPC controllers and solves the finite horizon optimal control problem once in a control period \cite{Zheng2012_13}, \cite{Vaccarini14} and \cite{Zheng2011_15}. In the DMPC with an iterative-based algorithm, each local MPC controller communicates with other neighboring agents several times per time-step \cite{Giselsson16}, \cite{Zheng2009_17} and \cite{Li2005_18}.

Also, they can be categorized based on the type of cost function, which is optimized per time-step into cooperative or non-cooperative algorithms. In the non-cooperative version of DMPC, each local MPC controller optimizes the cost function of its corresponding subsystem \cite{Farina19}, \cite{Dunbar20}, \cite{Maxim21} and \cite{Yuan22} . In its cooperative version, to obtain a better performance, each local controller optimizes a global cost function \cite{Giselsson16} and \cite{Zheng2018_23}. So, in this version, each MPC agent needs a network to exchange information with all other subsystems. In another version of cooperative DMPC, to provide a trade-off between performance and communication cost, a novel coordination strategy was proposed in \cite{Zheng2009_17} and \cite{Zhang2007_24}. Here each MPC agent minimizes the cost function of its own subsystem and the cost function of the subsystems directly impacts them.
Another classification of DMPC can be done in terms of the topology of the communication network into fully connected or partially connected networks. In the DMPC with a fully connected structure \cite{Li2005_18}, \cite{Stewart25}, \cite{Liu26} and \cite{Chen27} , exchange of information can be done among of all local controllers. In contrast, in DMPC with a partially connected structure \cite{Farina19}, \cite{Zhang2007_24}, \cite{Alessio28} and \cite{RawlingsPaper29}, it is required that send and receive of information are only done among neighboring subsystems \cite{Scattolini9}.

For a distributed implementation of MPC regulators, various approaches have been proposed in the literature. The most widely employed idea is to consider the interactions between subsystems as local disturbances and then to employ the local robust MPC (RMPC) controllers to handle them. \cite{Farina19} and \cite{BettiBook31} employed a non-cooperative, non-iterative DMPC for control of a system composed of linear discrete-time dynamically interconnected subsystems. In their proposed DMPC method, an RMPC has been employed as a local controller for each subsystem. The satisfaction of state and input constraints, guaranteeing convergence of the closed-loop system, obtaining the local control inputs of each subsystem without needing the dynamical models of the other subsystems and to be the limited transmission of information are the main traits of their proposed DMPC. Also, some realization issues for practical use of their proposed method, such as the automatic selection of some tuning parameters, the initialization of the algorithm, or its response to unexpected disturbances, have been discussed and solved in \cite{BettiPaper32}. The current research employs their proposed DMPC for control of distributed switched large-scale systems. For the reason of the existence of switch in the communication network of this kind of system, it is required to use a switched RMPC controller instead of a simple RMPC for each subsystem.

Recently, some researches have been applied the MPC to switched systems. A controller with a predictive framework has been proposed in \cite{Mhaskar2005_33} for the constrained stabilization of switched nonlinear systems with an apriori known switching signal. \cite{Mhaskar2008_34} extended this work for switched non-linear systems in which the switching times are unknown but only lie in a prior known interval. \cite{Muller35} presented an MPC for asymptotic stabilization of switched nonlinear systems whose switching times were a priori unknown but under restrictions of average dwell-time and detection fast enough. In \cite{Zhang36} a switched MPC (SwMPC) of a class of discrete-time switched linear systems with idea of mode-dependent dwell-time (MDT) of variable lengths was investigated. This research uses a SwMPC proposed recently in \cite{Danielson37}. By defining a new type of control invariant sets, called switch–robust control invariant (switch-RCI) sets, which are robust to unknown mode switching and under restrictions of minimum dwell-time and admissible mode transitions, it derived necessary and sufficient conditions for guaranteeing constraint satisfaction to control of constrained switched systems. The switch-RCI sets were used in the framework of a SwMPC to ensure constraint satisfaction in the presence of unknown mode switching with known minimum dwell-time.

Motivated by the above discussion, We employ a distributed SwMPC (DSwMPC) for the distributed switched constrained large-scale systems with input and state constraints.
\subsection{Switch and large-scale systems: a literature review} \label{SubSect1-1:SwSysRev}
The published works about the relation of the switch and large-scale systems can be categorized into two main groups. The first one includes those of researches in which the switch exists in the network topology of the system. One of the published studies on distributed control of distributed switched large-scale system is related to \cite{Zhang3}. They investigated sensor-network-based distributed control of networked control systems, in which the communication constraint and topology switching problems were addressed. Based on the Lyapunov direct method and the switched system approach, a sufficient condition was established to ensure the exponentially stability of the closed-loop system.
Also, Schiffer et al. \cite{Schiffer38} studied large-scale systems with the dynamic communication topology. In this research, first they derived a strict Lyapunov function for a nominal distributed averaging-based integral (DAI) -controlled power system model without communication uncertainties. Then, they extended this strict Lyapunov function to a common Lyapunov–Krasovskii functional to provide sufficient delay-dependent conditions for robust stability of a DAI-controlled power system with dynamic communication topology as well as heterogeneous constant and fast-varying delays.
The single published research in which the MPC was employed for the distributed large-scale systems with switched topology is related to Shi et al. \cite{Shi39}. They proposed a controller with a hierarchical structure for a class of distributed networked control systems with quantization and switching topology. An upper control layer receives the system-wide information and solves the MPC optimization problem in a centralized fashion, while the local controllers are implemented in a distributed way. Based on the switched control system theory and the Lyapunov direct method, the desired controller is designed by solving a constrained LMI optimization problem at each time step.

The second group includes the researches in which the switch has been inserted in the network topology of controllers to provide a reconfiguration capability. In this group, the topology of the applied system does not include an online switching. Baldivieso et al. \cite{Monasterios40} studied the nesting properties of feasible regions and robust positive invariant sets for different partitions of the large-scale system. It investigated how subsystems may be grouped together into coalitions. A coalition of subsystems is a non-empty subset of all subsystems. Their idea is that each coalition of subsystems operates and is controlled as a single entity. A coalitional controller replaces local subsystem controllers and may achieve better performance, albeit at a higher cost of complexity and communication. Also a coalitional control framework based on a switching MPC architecture aimed at large-scale dynamically-coupled systems was employed by Fele et al. \cite{Fele41}. Their proposed game theoretical framework provides a dynamic establishment of cooperation in the control of a multi-agent system. The obtained results showed that the reconfiguration capabilities provided to the system through the proposed framework were suited for fault-tolerance needs or plug-and-play settings.

A switching mechanism for communication between local MPC controllers for large-scale systems was employed in \cite{Nunez42}. Nunez et al. \cite{Nunez42} proposed a time-varying scheme for noncentralized MPC (NC-MPC) of large-scale systems. Their control strategy is implemented by clustering dynamically the local MPC controllers. In this strategy, several possible control structures for the communication between subsystems are considered and the hierarchical control system implements the one with the best performance according to a set of given objectives. Barreiro-Gomez et al. \cite{Barreiro-Gomez43} studied two fundamental components of the design of distributed optimization-based controllers for large-scale systems, i.e., system partitioning and distributed optimization algorithms. They combined the DMPC with the distributed partitioning algorithm with static and dynamical system partitioning. The obtained results of these two DMPC controllers demonstrated the effectiveness of both the density-dependent population game approach and the partitioning for a large-scale system.

Also, in some researches, the MPC is employed to handle plug and play operations in the structure of large-scale systems. Plug-and-play operation means that we exchange a subsystem in the overall system with another, possibly different, one or we add or remove a subsystem to the overall system. Lucia et al. \cite{Lucia44} proposed a DMPC to control coupled, possibly large-scale, systems for the case of plug-and-play operations. They explicitly consider the case in which a subsystem of the network is exchanged by a new one, but, the same physical interconnections. In their proposed method, when an exchange request is received, it is checked if the new subsystem can fulfill the contracts of the old subsystem, given the contracts of its future neighbors and its own dynamics and degrees of freedom. If this is possible, there exists no need to perform any redesign of the controllers. If the new subsystem cannot comply directly with the existing contracts, the controllers of all the neighbors can be virtually redesigned/tuned at the time which the request takes place. If the virtual redesign is infeasible, the exchange operation is rejected; otherwise it is performed.
\cite{Hou2021distributed} designed a DMPC  for networked systems where certain subsystems may be removed or inserted. In their proposed DMPC for this kind of reconfigurable systems, once the switch of topology is commanded, some additional optimization problems required to be solved in the related subsystems’ controllers to ensure the feasibility of removal or plugging-in operation.
\cite{Bai2020distributed}  investigated the DMPC of linear systems whose network topologies are changeable by the way of inserting new subsystems, disconnecting existing subsystems, or merely modifying the couplings between different subsystems. They firstly presented a distributed reconfiguration control strategy applying to any reconfigurable requirements. Then, based on the alternating direction method of multipliers (ADMM) algorithm, the way to redesign the reconfigured DMPC controller was provided and the iterative formulas employed in solving the reconfiguration optimization problem via ADMM algorithm were derived.

Based on categories mentioned above, in the current research, a switch exists in both the network topologies of controllers and system. The contribution of the paper is to apply the RSwMPC controllers in a distributed fashion on the distributed large-scale systems with switched topology, input and state constraints, in which the subsystems interact with each other by states and inputs. In comparison with \cite{Shi39}, this work considers a more general system that includes both input and state constraints and both input and state interactions. Also, the structure of the controller in \cite{Shi39}, is hierarchical and the MPC optimization problem has to be solved in a centralized fashion. So its computational complexity increases when the modes of switching signal and the number of the subsystems increase. In our proposed DSwMPC, the MPC optimization problems are locally solved for each controller, so the computational burden is considerably less than a centralized structure. Also, we already successfully applied a decentralized SwMPC (DeSwMPC) on distributed switched large-scale systems in \cite{Ahandani1}. In comparison with \cite{Ahandani1}, structure of the controller proposed in the current research is distributed and the interactions among subsystems in \cite{Ahandani1} were only based on states, whereas the proposed approach in the current research can handle states and input interactions.

The rest of the paper is organized as follows. In the next section, distributed switched large-scale systems are formulated. In Sect.~\ref{Sect3:DMPC}, the employed DMPC strategy is presented. Sect.~\ref{Sect4:SwMPC} contains the used MPC algorithm for switched constrained systems. The proposed switched DMPC is presented in Sect.~\ref{Sect5:DSwMPC}. The simulation results are presented and analyzed in Sect.~\ref{Sect6:Simulation}. Sect.~\ref{Sect7:Conclutions} concludes the paper.
\subsection{Notations and Definitions} \label{SubSect1-2:NotDef}
The Minkowski addition and subtraction of sets are denoted by the symbols $\oplus$  and $\ominus$ , respectively, and they are defined as follows: $C=A\oplus B=\{c=a+b: \hspace{0.1cm} for \hspace{0.1cm} all \hspace{0.1cm} a\in A, \hspace{0.1cm}b\in B \}$  and $C=A\ominus B=\{c:c+b\in A \hspace{0.1cm}, for \hspace{0.1cm} all \hspace{0.1cm}b\in B \}$ . A set ${\mathcal O}$  is a control invariant set for the system $x(t+1)=f(x,u)$ if there exists an admissible control law $u(x)\in {\mathcal U}$  such that for all $x\in {\mathcal O}$ , $f(x,u)\in {\mathcal O}$  and for all $t\geq 0$ . Also, ${\mathcal O}$  is positive invariant set for $x(t+1)=f(x)$  if for all $x\in {\mathcal O}$ , $f(x)\in {\mathcal O}$  and for all $t\geq 0$ .
A necessary and sufficient condition to be control invariance and as-well-as positive invariance is $ {\mathcal O}\subseteq Pre({\mathcal O})$  where $Pre({\mathcal S})$  is the predecessor operator that is the set of states   that can be directed to the target set ${\mathcal S}$  under the dynamics of the system  $x(t+1)=f(x,u)$ , i.e. $Pre({\mathcal S})=\{x|\exists u, f(x,u)\in {\mathcal S} \}$  without violating constraints. Also the set of $Pre^{k}({\mathcal S})\subseteq {\mathcal X}$  is the set of states $x\in {\mathcal X}$  that can be directed to the target set  ${\mathcal S}$ under the dynamics of the system in $k$ discrete-time instances without violating the constraints. The set $\Omega \subseteq {\mathcal R}^{n_{x}}$  is robust positively invariant (RPI) set for the system  $x(t+1)=f(x,w)$ where $w\in {\mathcal W} \subseteq {\mathcal R}^{n_{x}}$  is a disturbance vector, if $f(x,w)\subseteq \Omega$  for all $x\in \Omega$  and all $w\in {\mathcal W}$ and for all  $t\geq 0$, i.e. if only if $f(\Omega,{\mathcal W})\subseteq \Omega$. The RPI set $\underline{\Omega}$  is minimal if it is contained in every other RPI $\Omega$  verifies $\underline{\Omega} \subseteq \Omega$ .
\section{Problem Statement} \label{Sect2:Problem}
Large-scale systems have traditionally been characterized by large numbers of variables, the structure of interconnected subsystems, and other features that complicate the control models such as nonlinearities, time delays, and uncertainties. The decomposition of this kind of system into interconnections of a set of smaller-scale and more manageable subsystems allowed for implementing effective decentralization and coordination mechanisms \cite{Filip45}. This decomposition or partitioning is often performed for either computational or practical reasons.
In order to design a control structure based on robust control for large-scale systems, this research is used the modeling proposed in \cite{Li2016_6}. This paper considers the constrained discrete-time LTI distributed large-scale system,  , described by the following state space model:
\begin{eqnarray} \label{Eq1}
x(t+1)=Ax(t)+Bu(t) \\
x(t)\in {\mathcal X} , u(t)\in \hspace{0.05cm} {\mathcal U} \nonumber
\end{eqnarray}
where $x(t)\in  {\mathcal R}^{n_{x}}$  and $u(t)\in  {\mathcal R}^{n_{u}}$  are the state and input of system, respectively. ${\mathcal X}$  and  ${\mathcal U}$ are state and control constraint sets. It is assumed that the distributed large-scale system ${\mathcal S}$  can be decomposed of $M$  discrete-time linear subsystems ${\mathcal S_{i}}$ , $i\in  {\mathcal P=\{1, \ldots M} \}$  each of which is controlled by controllers  ${\mathcal C_{i}}$. Then, the subsystem  ${\mathcal S_{i}}$ can be expressed as
\begin{equation} \label{Eq2}
x_{i}(t+1)=A_{ii}x_{i}(t)+B_{ii}u_{i}(t)+\sum_{j\in {\mathcal N_{i}}} A_{ij}x_{j}(t) \\
           +\sum_{j\in {\mathcal N_{i}}} B_{ij}u_{j}(t)
\end{equation}
where  $x_{i}(t)\in  {\mathcal R}^{n_{x_{i}}}$  and $u_{i}(t)\in  {\mathcal R}^{n_{u_{i}}}$ are the state and input vectors of i-th subsystem, respectively, $A_{ii}\in {\mathcal R}^{{n_{x_{i}}}\times {n_{x_{i}}}}$ , $B_{ii}\in {\mathcal R}^{{n_{u_{i}}}\times {n_{u_{i}}}}$ , $A_{ij}\in {\mathcal R}^{{n_{x_{i}}}\times {n_{x_{j}}}}$  and $B_{ii}\in {\mathcal R}^{{n_{u_{i}}}\times {u_{x_{j}}}}$ . ${\mathcal N}_{i}$  is the set of neighbors of  ${\mathcal S}_{i}$ defined as: ${\mathcal N}_{i}=\{j\in {\mathcal P} : A_{ij}\neq 0 \hspace{0.1cm} and/or \hspace{0.1cm} B_{ij}\neq 0 , i\neq j \}$ .
It is said that subsystem ${\mathcal S}_{j}$  is a dynamic neighbor of subsystem ${\mathcal S}_{i}$  if only if at least one of the matrices $A_{ij}$ or $B_{ij}$  is nonzero, so ${\mathcal N}_{i}$  will be non-empty. The subsystem  ${\mathcal S}_{i}$ includes the local (uncoupled) state and control constraints
\begin{equation} \label{Eq3}
x_{i}(t)\in \hspace{0.05cm} {\mathcal X}_{i}\subseteq {\mathcal R}^{n_{x_i}} , \hspace{0.1cm} u_{i}(t)\in \hspace{0.05cm} {\mathcal U}_{i}\subseteq {\mathcal R}^{n_{u_i}}
\end{equation}
So
\begin{equation} \label{Eq4}
x(t)=[x_{1}^{T}(t), x_{2}^{T}(t), \ldots , x_{M}^{T}(t)]\in {\mathcal X}\subseteq {\mathcal R}^{n_{x}}
\end{equation}
\begin{equation} \label{Eq5}
u(t)=[u_{1}^{T}(t), u_{2}^{T}(t), \ldots , u_{M}^{T}(t)]\in {\mathcal U}\subseteq {\mathcal R}^{n_{u}}
\end{equation}
Also, $n_x=\sum_{i\in {\mathcal P}} n_{x_i}$ , $n_u=\sum_{i\in {\mathcal P}} n_{u_i}$ ,${\mathcal X}=\Pi_{i=1}^{M} {\mathcal X}_{i}$ and ${\mathcal U}=\Pi_{i=1}^{M} {\mathcal U}_{i}$. Also the interacted subsystems can be subject to global (coupled) constraints described in collective form by
\begin{equation} \label{Eq6}
\sum_{i=1}^{M} (\psi_{i}^{x}x_{i}(t)+\psi_{i}^{u}u_{i}(t))\leq 1_p
\end{equation}
where the matrices $\psi_{i}^{x}\in {\mathcal R}^{p\times {n_{x_{i}}}} $  and $\psi_{i}^{u}\in {\mathcal R}^{p\times {n_{u_{i}}}} $  define the global constraint for subsystem ${\mathcal S}_{i}$ and  $1_p$ is a  $p$~-vector of all ones. By definition of the interaction vector $w_i$ , named as perturbation or disturbance vector, Eq.~(\ref{Eq2}) is rewritten as follows:
\begin{equation} \label{Eq7}
x_{i}(t+1)=A_{ii}x_{i}(t)+B_{ii}u_{i}(t)+w_{i}(t)
\end{equation}
where
\begin{equation} \label{Eq8}
w_{i}(t)= \sum_{j\in {\mathcal N_{i}}} A_{ij}x_{j}(t)+\sum_{j\in {\mathcal N_{i}}} B_{ij}u_{j}(t)
\end{equation}
In the distributed switched large-scale systems, a switching signal changes the network topology over time. So this kind of system is described by the following state space model:
\begin{eqnarray} \label{Eq9}
x(t+1)=A^{\sigma(t)}x(t)+B^{\sigma(t)}u(t) \\
x(t)\in {\mathcal X}^{\sigma(t)} , u(t)\in {\mathcal U}^{\sigma(t)} \nonumber
\end{eqnarray}
where the switching sequence $\sigma:\mathcal{N}\rightarrow \mathcal{I}$  is a known or unknown exogenous input that switches network topology between a finite number of modes $ \mathcal{I}\subset \mathcal{N}$ . From the viewpoint of the dynamic of the $i$~-the subsystem, due to switching signal varies the interactions among subsystems, it affects only the interaction vector $w_i$ . In other words, a switch can only change the set of neighbors of ${\mathcal S}_{i}$ , i.e.  ${\mathcal N}_{i}$. So the dynamic equation of subsystem  ${\mathcal S}_{i}$ in such a system can be expressed as
\begin{equation} \label{Eq10}
x_{i}(t+1)=A_{ii}x_{i}(t)+B_{ii}u_{i}(t)+w_{i}^{\sigma  (t)}(t)
\end{equation}
where
\begin{equation} \label{Eq11}
w_{i}^{\sigma  (t)}(t)= \sum_{j\in {\mathcal N_{i}^{\sigma  (t)}}} A_{ij}x_{j}(t)+\sum_{j\in {\mathcal N_{i}^{\sigma  (t)}}} B_{ij}u_{j}(t)
\end{equation}
Eqs.~(\ref{Eq10}) and (\ref{Eq11}) clearly show that the local constraints of subsystems have not been affected by switch in network topology but the global or interaction constraints can be varied by a switching signal \cite{Ahandani1}.
So, the system employed in this research to be controlled is a distributed switched large-scale system described by the state space model of (\ref{Eq9}). This system includes $M$  coupled subsystems with dynamics shown in (\ref{Eq10}) and (\ref{Eq11}). In this networked modeling with time-varying topology, the role of switching signal $\sigma  (t)$  is that it can provide a possibility to vary the set of neighbors of subsystems.   This paper, based on this modeling, presents a distributed control structure to be applied on such systems by considering the switched interaction term among the subsystems as disturbances to be rejected.
\section{Distributed robust tube-based MPC} \label{Sect3:DMPC}
This section describes the distributed robust tube-based MPC proposed in \cite{Farina19}, \cite{BettiBook31} and \cite{BettiPaper32} for control of large-scale constrained discrete-time linear systems consisting dynamically coupled subsystems. This DMPC in combination with SwMPC will be adopted to switch in the network topology of the system (see Sect. 5).

The DMPC presented in \cite{Farina19}, \cite{BettiBook31} and \cite{BettiPaper32} guarantees stability and convergence properties of the system (\ref{Eq1}) and constraints satisfaction under mild assumptions. The dynamic equations of subsystems,   in (\ref{Eq2}), can be subjected to local and global state and control constraints. In their proposed DMPC, at each time instant, the subsystem  ${\mathcal S}_{i}$ sends information about its future state $\tilde{x}_{i}$  and input $\tilde{u}_{i}$  reference trajectories, calculated based on the solution of optimal control problem in the previous time instant, to its interconnecting subsystems. The DMPC solves the optimal control problem of current time instant as its actual trajectories $x_{i}$  and $u_{i}$  lie within certain regions in the neighborhood of the reference ones. The DMPC limits the error between reference trajectories, i.e.  $\tilde{x}_{i}$  and input $\tilde{u}_{i}$ , and predictive states calculated in the current sampling time within certain regions by adding some new constraints to the optimal control problem of each subsystem. So the DMPC considers the differences  $x_{i}-\hat{x}_{i}$  and input $u_{i}-\hat{u}_{i}$  as unknown bounded disturbances and it employs a tube-based RMPC controller motivated by the discussion in \cite{Mayne46} for each subsystem to reject them.

Consider the dynamic of a subsystem  ${\mathcal S}_{i}$ as (\ref{Eq2}) subjects to constraints of (\ref{Eq3}) where ${\mathcal X}_{i}$  and  ${\mathcal U}_{i}$ are convex neighborhoods of the origin. Also, for the system (\ref{Eq2}), it can be considered the global static constraints as follows
\begin{equation} \label{Eq12}
H_{s}(x(t),u(t))\leq 0
\end{equation}
where $s=1,\ldots , n_c$ . The subsystem ${\mathcal S}_{i}$  is subjected to constraint $H_{s}$  if $x_i$  and/or $u_i$   are arguments of $H_{s}$ , while ${\mathcal D}_{i}=\{s \hspace{0.1cm}\in \hspace{0.1cm} \{1,\ldots , n_c\}: H_c \hspace{0.17cm} is \hspace{0.17cm} constraint \hspace{0.17cm} on \hspace{0.2cm} i\}$  denotes the set of constraints on ${\mathcal S}_{i}$ . The subsystem ${\mathcal S}_{j}$  is a constraint neighbor of subsystem ${\mathcal S}_{i}$  if there exists  $\hspace{0.1cm}\bar{s}\in {\mathcal D}_{i}$ such that $x_j$  and/or $u_j$  are arguments of $H_{\bar{s}}$  , while ${\mathcal H}_{i}$  denotes the set of the constraint neighbors of ${\mathcal S}_{i}$  \cite{Farina19}, \cite{BettiBook31} and \cite{BettiPaper32}. In general, ${\mathcal S}_{j}$  is called a neighbor of ${\mathcal S}_{i}$  if  $j \in {\mathcal P}_{i}$ where ${\mathcal P}_{i}={\mathcal N}_{i}\cup {\mathcal H}_{i}$ .
Concerning system (\ref{Eq1}) and its partition, the following primary assumption on decentralized stabilizability is introduced.

\textbf{Assumption 1}.

(i) The matrices $F_{ii}=A_{ii}+B_{ii}K_i$ ,  $i\in {\mathcal P}_{i}$ are Schur.

(ii) The matrix $F=A+BK$  is Schur.

where $K=diag\{K_{1} ,\ldots , K_{M}\}$ .

Let $\tilde{x}_i(t+k)$  and  $\tilde{u}_i(t+k)$ are future state and input reference trajectories transmitted by ${\mathcal S}_{i}$  to its neighbor subsystems. In DMPC some added constraints to finite-horizon optimal control problem ensure that real trajectories of each subsystem lie in the specified time-invariant neighborhoods of their reference trajectories, i.e., $x_i(t)\in \tilde{x}(t)+ \mathcal{E}_i$  and $u_i(t)\in\tilde{u}(t)+ \mathcal{E}_i^u$ , where $0\in \mathcal{E}_i$  and $0\in \mathcal{E}_i^u$ . So the dynamic equation of the subsystem ${\mathcal S}_{i}$  in (\ref{Eq2}) can be expressed as
\begin{eqnarray} \label{Eq13}
x_{i}(t+1)=A_{ii}x_{i}(t)+B_{ii}u_{i}(t)+\sum_{j\in {\mathcal N_{i}}} A_{ij}\tilde{x}_{j}(t)
           +\sum_{j\in {\mathcal N_{i}}} B_{ij}\tilde{u}_{j}(t)+w_i^\varepsilon (t)
\end{eqnarray}
where
\begin{equation} \label{Eq14}
w_i^\varepsilon (t)=\sum_{j\in {\mathcal N_{i}}} A_{ij}({x}_{j}(t)-\tilde{x}_{j}(t) )
           +\sum_{j\in {\mathcal N_{i}}} B_{ij}({u}_{j}(t)-\tilde{u}_{j}(t) )\in {\mathcal W_{i}}
\end{equation}
and ${\mathcal W_{i}}=\bigoplus_{j\in {\mathcal N_{i}}} (A_{ij}\mathcal{E}_j+B_{ij}\mathcal{E}_j^u)$ . In DMPC, a local robust MPC is employed for each subsystem with the dynamic of (\ref{Eq13}), where the term $\sum_{j\in {\mathcal N_{i}}} A_{ij}\tilde{x}_{j}(t+k)+\sum_{j\in {\mathcal N_{i}}} B_{ij}\tilde{u}_{j}(t+k)$  is a prior known input over the prediction horizon $k=0,\ldots N_i-1$  and $w_i^\varepsilon (t)$  is a bounded disturbance to be rejected. In order to produce predictions, a nominal model for subsystem ${\mathcal S_{i}}$  is obtained by removing the disturbance term of $w_i^\varepsilon (t)$  in (\ref{Eq13}):
\begin{eqnarray} \label{Eq15}
x_{i}(t+1)=A_{ii}x_{i}(t)+B_{ii}u_{i}(t)+\sum_{j\in {\mathcal N_{i}}} A_{ij}\tilde{x}_{j}(t)
 +\sum_{j\in {\mathcal N_{i}}} B_{ij}\tilde{u}_{j}(t)
\end{eqnarray}
So the local controller  ${\mathcal C_{i}}$ when the current state of subsystem ${\mathcal S_{i}}$  is $x_i(t)$ , is defined in the form of an implicit MPC law as follows:
\begin{equation} \label{Eq16}
u_i(t)=\hat{u}_i(t)+K_i(x_i(t)-\hat{x}_i(t))
\end{equation}
where $K_i\in{\mathcal R}^{n_{u_i}\times n_{x_i}}$ , $i\in{\mathcal P}_i$  must be chosen to satisfy Assumption 1. A local MPC controller is employed to obtain two unknown variables of $\hat{u}_i(t)$  and $\hat{x}_i(t)$ .

Given (\ref{Eq13}), (\ref{Eq15}) and (\ref{Eq16}), and since $F_{ii}=A_{ii}+B_{ii}K_i$  is Schur and $w_i^\varepsilon \in{\mathcal W}_i$  is bounded, they defined the nonempty RPI sets  , ${\mathcal Z}_i$ (see \cite{Farina19}). Based on ${\mathcal Z}_i$ , two new sets, neighborhoods of origin,  $\Delta \mathcal{E}_i$ and $\Delta{\mathcal U}_i$  for $i=1,\ldots,M$  are defined in a way that $\Delta \mathcal{E}_i\oplus {\mathcal Z}_i \subseteq \mathcal{E}_i$  and $\Delta \mathcal{U}_i\oplus K_i{\mathcal Z}_i \subseteq \mathcal{E}_i^u$ , respectively \cite{Mayne46}.

By assuming that each subsystem $\mathcal{S}_i$  knows future reference trajectories of its neighbors over the entire prediction horizon, i.e. $\tilde{x}_j(t+k)$  and  $\tilde{u}_j(t+k)$, $k=0,\ldots, N_j-1$  and $j\in\mathcal{N}_i\cup\mathcal{H}_i\cup\{i\}$ , the unknown variables of (\ref{Eq16}) are computed by online solving the MPC local performance index of subsystem $\mathcal{S}_i$  described in (\ref{Eq19}) at each sampling time $t$
\begin{subequations} \label{Eq19}
\begin{equation} \label{Eq19a}
\textbf{min}_{\substack {\hat{x}_i(0)\\
\hat{u}_i(0:N_i-1)}}
\sum_{k=0}^{N_i-1}(\|\hat{x}_i(k|t)\|^2_{Q_i^0}+\|\hat{u}_i(k|t)\|^2_{R_i^0})+ \|\hat{x}_i(N_i|t)\|^2_{P_i^0}
\end{equation}
subject to (\ref{Eq13}),
\begin{equation} \label{Eq19b}
x_i(t)-\hat{x}_i(t)\in\mathcal{Z}_i \hspace{1cm}
\end{equation}
\begin{equation} \label{Eq19c}
\hat{x}_i(t+k|t)-\tilde{x}_i(t+k|t)\in\Delta \mathcal{E}_i \hspace{0.8cm} k=0,\ldots,N_i-1
\end{equation}
\begin{equation} \label{Eq19d}
\hat{u}_i(t+k|t)-\tilde{u}_i(t+k|t)\in\Delta \mathcal{U}_i \hspace{0.7cm} k=0,\ldots,N_i-1
\end{equation}
\begin{equation} \label{Eq19e}
\hat{x}_i(t+k|t)\in \hat{\mathcal{X}}_i \hspace{0.1cm}, \hspace{0.1cm}
\hat{u}_i(t+k|t)\in \hat{\mathcal{U}}_i \hspace{0.1cm}  \hspace{0.6cm} k=0,\ldots,N_i-1
\end{equation}
\begin{equation} \label{Eq19f}
\hat{x}_i(t+N_i|t)\in \hat{\mathcal{X}}_{f_i}
\end{equation}
\end{subequations}
where $N_i\in \mathcal{N}$  is the prediction horizon, $l_i(\hat{x}_i(k|t),\hat{u}_i(k|t))=(\|\hat{x}_i(k|t)\|^2_{Q_i^0}+\|\hat{u}_i(k|t)\|^2_{R_i^0})$  is stage cost, $V_{f_i}(\hat{x}_i(t+N_i|t))=\|\hat{x}_i(N_i|t)\|^2_{P_i^0}$  is terminal cost and $\hat{\mathcal{X}}_{f_i}$  is terminal region.

The local state and input constraints for the nominal system (\ref{Eq15}) are $\hat{\mathcal{X}}_i$  and $\hat{\mathcal{U}}_i$ , respectively. These constraints, used in (\ref{Eq19e}), can be obtained after computing the RPI set ${\mathcal{Z}}_i$  as
\begin{equation} \label{Eq20}
\hat{\mathcal{X}}_i={\mathcal{X}}_i\ominus {\mathcal{Z}}_i
\end{equation}
\begin{equation} \label{Eq21}
\hat{\mathcal{U}}_i={\mathcal{U}}_i\ominus K_i\mathcal{Z}_i
\end{equation}

The positive-definite stage and terminal matrices in (\ref{Eq19}), i.e. $Q_i^0$ , $R_i^0$  and $P_i^0$ , are chosen based on algorithms discussed in \cite{BettiBook31} to ensure the stability of MPC. Also, other design parameters employed in DMPC i.e. the sets $\mathcal{Z}_i$ , $\Delta \mathcal{E}_i$  and $\Delta \mathcal{E}_i^u$ , $\hat{\mathcal{X}}_{f_i}$ and the gain matrices $K_i$  are calculated as proposed in \cite{BettiBook31}.

\section{Switched model-based predictive control} \label{Sect4:SwMPC}
Danielson et al. \cite{Danielson37} studied the control of switched constrained linear systems by defining a new kind of control invariant sets for switched constrained systems robust to unknown mode switching, named switch-RCI sets. These switch-RCI sets are used to derive necessary and sufficient conditions for the existence of a control-law to ensure constraint satisfaction in the presence of unknown mode switching with known minimum dwell-time. The switch-RCI sets are also employed to design a recursively feasible MPC that enforces closed-loop constraint satisfaction for switched constrained systems. This section briefly describes the SwMPC proposed by \cite{Danielson37}.

Consider a switched constrained system described as follows:
\begin{eqnarray} \label{Eq22}
x(t+1)=f_{\sigma(t)}(x(t),u(t)) \\
x(t)\in {\mathcal X}_{\sigma(t)} , u(t)\in {\mathcal U}_{\sigma(t)} \nonumber
\end{eqnarray}
where $x(t)\in {\mathcal R}^{n_x}$  is the state and $u(t)\in {\mathcal R}^{n^{\sigma(t)}_u}$  is the input. The switching sequence $\sigma:\mathcal N \rightarrow\mathcal I $  is an unknown exogenous input that switches the dynamics $f_i:{\mathcal R}^{n_x}\times{\mathcal R}^{n_u}\rightarrow {\mathcal R}^{n_x}$  and the constraint sets  $\mathcal X\in {\mathcal R}^{n_x}$  and $\mathcal U\in {\mathcal R}^{n^i_u}$  between a finite number of modes  $\mathcal I \subset\mathcal N $. The number of inputs $n^i_u$  may depend on the mode $i\in\mathcal I$ . \cite{Danielson37} implemented two common constraints on the switching sequence, i.e. dwell-time and mode transition restrictions. The set of switching signals $\sigma$  that satisfies these restrictions is described as follows:
\begin{eqnarray} \label{Eq25}
\sum(d,\mathcal{G})=\{\sigma:\mathcal N \rightarrow\mathcal I: dwell_i(\sigma)\geq d_i ,
  (\sigma(\tau_{s}),\sigma(\tau_{s+1}))\in E,\forall s\in \mathcal{N}\}
\end{eqnarray}
where
\begin{eqnarray} \label{Eq25new}
dwell_i(\sigma)=min\{\tau_{s+1}-\tau_{s}: \sigma(\tau_{s})=i, s\in \mathcal{N}\}
\end{eqnarray}
and $dwell_i(\sigma)$ is dwell-time of a mode  $i\in\mathcal I$, $\tau_s\in\mathcal{N}$ are the discrete-times at which the mode changes $\sigma(\tau_s)\neq\sigma(\tau_s-1)$ and $\mathcal{G}=(\mathcal{I},E)$  is a directed graph in which the graph nodes $\mathcal{I}$  denote the modes of the switched system (\ref{Eq22}) and each directed edge $(i,j)\in E$  indicates that a switch from mode  $\sigma(\tau_s)=i$ to mode $\sigma(\tau_{s+1})=j$  is allowed. For brevity, we will use the short-hand $\sum=\sum(d,\mathcal{G})$  later.

The system (\ref{Eq22}) has a property of the time-varying depend on the remaining amount of dwell-time denoted by $\delta(t)$ , where
\begin{equation} \label{Eq24}
\delta(t+1)= \left \{
\begin{aligned}
    &max\{\delta(t)-1,0\} && \text{if}\ \sigma(t+1)=\sigma(t) \\
    &d_{\sigma(t+1)},      && \text{otherwise}
\end{aligned} \right.
\end{equation}
The objective of \cite{Danielson37} is to determine whether there exists a control-law $u=k(x,\sigma,\delta)$  that ensures constraints satisfaction $u(t)\in\mathcal{U}_{\sigma(t)}$  and $x(t)\in\mathcal{X}_{\sigma(t)}$  for all allowable switching signals $\sigma(t)\in \sum$\hspace{0.05cm}  and subsequent times $t\geq 0$  provided that initial states, modes, and remaining dwell-times are $x_0$ ,  $\sigma_0$ and $\delta_0$ , respectively.

They defined a new class of control invariant sets for switched constrained systems (\ref{Eq22}) named switch-RCI sets. Switch-RCI sets  $\{\mathcal{C}_i\}_{i\in\mathcal I}$ are constraint admissible control invariant sets that are robust to unknown mode switching provided that switching sequences are contained in (\ref{Eq25}). They are defined in Definition 1.

\textbf{Definition 1} (Switch-RCI sets).
A collection of sets  $\mathcal{C}_i\subseteq \mathcal{X}_i$ for $i\in\mathcal I$  is named switch-RCI if they are control invariant, i.e. $\mathcal{C}_i\subseteq Pre_i^1(\mathcal{C}_i)$  , for each mode $i\in\mathcal I$  and mutually reachable $\mathcal{C}_i\subseteq Pre_j^{d_j}(\mathcal{C}_j)\subseteq \mathcal{X}_j$  within the dwell-time $d_j$  for each allowable mode transition  $(i,j)\in E$.

Then they by defining maximal switch-RCI sets in Definition 2, combined the sufficient and necessary conditions for the existence of a control-law to ensure constraint satisfaction.

\textbf{Definition 2} (Maximal Switch-RCI Sets).
A collection of sets  $\{\mathcal{C}_i\}_{i\in\mathcal I}^{\infty}$ is called maximal switch-RCI if they are switch-RCI and, for any group of switch-RCI sets $\{\mathcal{C}_i\}_{i\in\mathcal I}$ , we have $\mathcal{C}_i\subseteq\mathcal{C}_i^{\infty}$  for each mode  $i\in\mathcal I$.

\textbf{Theorem 1.} \label{Th1}
  A control-law  can be defined as  $k(x,\sigma,\delta)$ to guarantee constraint satisfaction for all times $t\in\mathcal{N}$  and for all allowable switching signals $\sigma\in \sum$  if and only if the initial state $x(t_0)$ , mode $\sigma(t_0)$ , and remaining dwell-time $\delta(t_0)$  are contained in the following initial conditions set for the maximal switch-RCI sets $\{\mathcal{C}_i\}_{i\in\mathcal I}^{\infty}$ :
\begin{equation} \label{Eq26}
IC=\{(x(t_0),\sigma(t_0),\delta(t_0)):(x(t_0)\in Pre_{\sigma(t_0)}^{\delta(t_0)}(\mathcal{C}_{\sigma(t_0)}) \}
\end{equation}
  \label{Thm1}

Pseudocode of Fig.~\ref{Fig1} computes the maximal switch-RCI sets $\{\mathcal{C}_i\}_{i\in\mathcal I}^{\infty}$  \cite{Danielson37}.

\cite{Danielson37} employed the maximal switch-RCI sets $\{\mathcal{C}_i\}_{i\in\mathcal I}^{\infty}$  to obtain terminal constraint set in the SwMPC to generate control inputs that satisfy state and input constraints. They formulated the SwMPC where (i) the prediction horizon  is longer than the dwell-time of each mode $i\in\mathcal I$  (i.e. $N\geq d_i$), named long-horizon SwMPC, (ii) the prediction horizon is shorter than the dwell-time of each mode $i\in\mathcal I$  (i.e. $N\leq d_i$), named short-horizon SwMPC. In this section, we concentrate on long-horizon SwMPC. The following proposed relations can be extended for short-horizon SwMPC with some modifications based on \cite{Danielson37}.
\begin{figure}
  \begin{algorithmic}[1]
\For{each mode $i\in\mathcal I$}
\State $\Omega_i^0=\mathcal{X}_i$
\EndFor
\Repeat
  \For{each mode $i\in\mathcal I$}
  \State update sets \\
          \hspace{1.2cm} $\Omega_i^{k+1}=\Omega_i^{k}\bigcap Pre_i(\Omega_i^{k})\bigcap_{(i,j)\in E} Pre_j^{d_j}(\Omega_j^{k}) $
  \EndFor
\Until{$\Omega_i^{k+1}=\Omega_i^{k} $ for all $i\in\mathcal I$}
\\ $\mathcal{C}_i^{\infty}=\Omega_i^{k} $ for all $i\in\mathcal I$
\end{algorithmic}
\caption{Pseudocode for computation of maximal switch-RCI sets \cite{Danielson37}.}
\label{Fig1}
\end{figure}

For the long-horizon SwMPC, the control input $u(t)$  is computed by solving the following open-loop constrained finite horizon optimal control problem:
\begin{subequations} \label{Eq27}
\begin{equation} \label{Eq27a}
\textbf{min}_{u(0:N-1)}
\sum_{k=0}^{N-1} l^{\sigma(0|t)}(x(k|t),u(k|t))+V^{\sigma(0|t)}_{f}(x(N|t))
\end{equation}
\begin{equation} \label{Eq27b}
x(t+k+1|t)=f_{{\sigma(0|t)}}(x(t+k|t),u(t+k|t)) \hspace{0.3cm} k=0,\ldots,N-1
\end{equation}
\begin{equation} \label{Eq27c}
x(t+k|t)\in {\mathcal{X}}^{\sigma(0|t)} \hspace{0.1cm}, \hspace{0.1cm}
u(t+k|t)\in {\mathcal{U}}^{\sigma(0|t)} \hspace{0.6cm}  k=0,\ldots,N-1
\end{equation}
\begin{equation} \label{Eq27d}
x(t+k|t)\in \mathcal{T}^{\sigma(0|t)} \hspace{0.6cm}  for \hspace{0.2cm} k\geq \delta(0|t)
\end{equation}
\end{subequations}
in which $x(0|t)=x(t)$  is the initial condition of the switched system defined in (\ref{Eq22}), $x(t+k|t)$  is the predicted state trajectory of the system driven by $u(t+k|t)$  over the prediction horizon $N\geq d_i$ , $\sigma(0|t)=\sigma(t)$  is the current mode of the system, $\delta(0|t)=\delta(t)$ is the remaining dwell-time, and  $\mathcal{T}^{\sigma(0|t)}$ is the switch-RCI set. By solving (\ref{Eq27}), the obtained control input guarantees the constraints satisfaction and the terminal cost, $V^{\sigma(0|t)}_{f}(\cdot)$ , and the stage cost, $l^{\sigma(0|t)}(\cdot,\cdot)$ , are assigned to satisfy secondary control objectives such as stability or reference tracking for the individual modes. In solving optimization problem (\ref{Eq27}), it is assumed that the mode  $\sigma(t)\in\mathcal I$  is constant $\sigma(k|t)=\sigma(0|t)$  over the entire horizon $k=0,\ldots,N-1$ .
The control input is the first element  $u^*(0|t)$ of the optimal open-loop input sequence $u^*(0|t)$,..., $u^*(N-1|t)$ i.e.,
\begin{equation} \label{Eq28}
u(t)=u^*(0|t)
\end{equation}
In contrast to traditional MPC, due to being depend on the current state $x(t)$ , mode  $\sigma(t)$, and the remaining dwell-time  $\delta(t)$, the controller (\ref{Eq28}) is mode-varying and time-varying \cite{Danielson37}. Also for more details about stability and the recursively feasible of (\ref{Eq27}) see \cite{Danielson37}.

\section{Distributed switched model-based predictive control} \label{Sect5:DSwMPC}
In order to adopt the DMPC of Sec. 3 for the large-scale systems with switched topology, this section presents the DSwMPC. In this method, the local RSwMPC controllers are used to handle the existence of switch in the network topology, instead of employing the RMPC regulators. The DSwMPC preserves all characteristics of DMPC such as asymptotic stability of the closed-loop system and constraints satisfaction. Because of employing the switch-RCI sets, the DSwMPC is robust to unknown mode switching. In other words, the DSwMPC is robust to unknown switching times and unknown next topologies, but as mentioned for SwMPC, it is assumed that a minimum dwell-time and allowable mode transitions are known in prior. Restriction of mode transitions for the distributed switched large-scale systems means that it is not necessary to know neighbors of a subsystem in prior but a set of allowable neighborhoods has to be known in prior.
Let the dynamic equation of a distributed switched large-scale system, $\mathcal{S}$ , as defined in (\ref{Eq9}). Suppose $\sigma(t)$  is an unknown switching signal that changes the network topology. Based on modeling proposed in (\ref{Eq10}), (\ref{Eq11}) and (\ref{Eq13}) for dynamic of $i$~-th subsystem, dynamic equation of subsystem $\mathcal{S}_i$  can be expressed as:
\begin{eqnarray} \label{Eq30}
&x_{i}(t+1)=A_{ii}x_{i}(t)+B_{ii}u_{i}(t)+\sum_{j\in {\mathcal N_i^{\sigma(t)}}} A_{ij}\tilde{x}_{j}(t)          +\sum_{j\in {\mathcal N_i^{\sigma(t)}}} B_{ij}\tilde{u}_{j}(t)+w_i^{\varepsilon, \sigma(t)} (t) \\
&x(t)\in {\mathcal X}_{\sigma(t)} , u(t)\in {\mathcal U}_{\sigma(t)}  \nonumber
\end{eqnarray}
where
\begin{eqnarray} \label{Eq31}
w_i^{\varepsilon, \sigma(t)} (t)=\sum_{j\in {\mathcal N_{i}}} A_{ij}({x}_{j}(t)-\tilde{x}_{j}(t) )
           +\sum_{j\in {\mathcal N_{i}}} B_{ij}({u}_{j}(t)-\tilde{u}_{j}(t) )\in {\mathcal W_i^{\sigma(t)}}
\end{eqnarray}
where $w_i^{\varepsilon, \sigma(t)} (t)\in\mathcal{W}_i^\sigma(t)=\bigoplus_{j\in {\mathcal N}_i^{\sigma(t)}} (A_{ij}\mathcal{E}_j+B_{ij}\mathcal{E}_j^u)$ . Observe that according to (\ref{Eq30}) and (\ref{Eq31}), the effect of a switch in the network topology appears in the terms of future state and input reference trajectories and also the term of additive disturbance in dynamic equation of $\mathcal{S}_i$ . Also, we know that when a switch occurs, due to the new possible neighborhood set, $\mathcal{Z}_i$ ,$i\in {\mathcal P}$ ,  must be updated based on the new network topology. On the other side, based on (\ref{Eq19b}), (\ref{Eq19c}), (\ref{Eq19d}), (\ref{Eq20}) and (\ref{Eq21}), $\mathcal{S}_i$  is utilized for computation of feasible regions of $\hat{x}_i(0)$ , $\hat{x}_i(t)$ , $\hat{u}_i(t)$  and also tightened constraints  $\hat{\mathcal{X}}_i$ and $\hat{\mathcal{U}}_i$ . So a switch in the network topology affects both the dynamic equation and constraints of the nominal subsystem in the structure of the proposed DSwMPC and therefore its effect appears in the optimization problem of (\ref{Eq19}). From the model defined in (\ref{Eq30}) for the dynamic equation of subsystem $\mathcal{S}_i$ , its corresponding nominal subsystem is considered as follows:
\begin{eqnarray} \label{Eq32}
&\hat{x}_{i}(t+1)=A_{ii}\hat{x}_{i}(t)+B_{ii}\hat{u}_{i}(t)+\sum_{j\in {\mathcal N_i^{\sigma(t)})}} A_{ij}\tilde{x}_{j}(t)+\sum_{j\in {\mathcal N_i^{\sigma(t)}}} B_{ij}\tilde{u}_{j}(t) \nonumber \\
&\hat{x}(t)\in\hspace{0.1cm} \hat{{\mathcal X}}_i^{\sigma(t)} , \hat{u}(t)\in \hat{{\mathcal U}}_i^{\sigma(t)}
\end{eqnarray}
where
\begin{equation} \label{Eq33}
\hat{{\mathcal X}}_i^{\sigma(t)}={\mathcal X}_i \ominus \hat{{\mathcal Z}}_i^{\sigma(t)}
\end{equation}
\begin{equation} \label{Eq34}
\hat{{\mathcal U}}_i^{\sigma(t)}={\mathcal U}_i \ominus K_i^{\sigma(t)}\hat{{\mathcal Z}}_i^{\sigma(t)}
\end{equation}
and $\hat{{\mathcal Z}}_i^{\sigma(t)}$  is the RPI set of ${\mathcal S}_i$  under switching signal of $\sigma(t)$ .

In the DMPC proposed in \cite{Farina19}, the controller ${\mathcal C}_i$  by assumption that the current state of ${\mathcal S}_i$  is $x_i$ , is obtained as proposed in (\ref{Eq16}) where two unknown variables of $\hat{x}_i(t)$  and $\hat{u}_i(t)$   are calculated using the MPC controller for the model defined in (\ref{Eq32}). The variables $\hat{x}_i(t)$  and $\hat{u}_i(t)$ are equal to optimization parameters of $\hat{x}_i(0)$  and $\hat{u}_i(0)$ , respectively. These parameters, for the long-horizon MPC, are obtained by optimizing the following MPC performance index:
\begin{subequations} \label{Eq35}
\begin{equation} \label{Eq35a}
\textbf{min}_{\substack {\hat{x}_i(0)\\
\hat{u}_i(0:N_i-1)}}
\sum_{k=0}^{N_i-1}(\|\hat{x}_i(k|t)\|^2_{Q_i^0}+\|\hat{u}_i(k|t)\|^2_{R_i^0})+ \|\hat{x}_i(N_i|t)\|^2_{P_i^0}
\end{equation}
\begin{eqnarray} \label{Eq35b}
\hat{x}_{i}(t+1)=A_{ii}\hat{x}_{i}(t)+B_{ii}\hat{u}_{i}(t)+\sum_{j\in {\mathcal N_i^{\sigma(t)})}} A_{ij}\tilde{x}_{j}(t) \\
           +\sum_{j\in {\mathcal N_i^{\sigma(t)}}} B_{ij}\tilde{u}_{j}(t) \hspace{0.5cm} k=0,\ldots,N_i-1 \nonumber
\end{eqnarray}
\begin{equation} \label{Eq35c}
x_i(t)-\hat{x}_i(t)\in\mathcal{Z}_i^{\sigma(t)} \hspace{1cm}
\end{equation}
\begin{equation} \label{Eq35d}
\hat{x}_i(t+k|t)-\tilde{x}_i(t+k|t)\in\Delta \mathcal{E}_i^{\sigma(t)} \hspace{0.8cm} k=0,\ldots,N_i-1
\end{equation}
\begin{equation} \label{Eq35e}
\hat{u}_i(t+k|t)-\tilde{u}_i(t+k|t)\in\Delta \mathcal{U}_i^{\sigma(t)} \hspace{0.7cm} k=0,\ldots,N_i-1
\end{equation}
\begin{equation} \label{Eq35f}
\hat{x}_i(t+k|t)\in \hat{\mathcal{X}}_i^{\sigma(t)} \hspace{0.1cm}, \hspace{0.1cm}
\hat{u}_i(t+k|t)\in \hat{\mathcal{U}}_i^{\sigma(t)} \hspace{0.1cm}  \hspace{0.6cm} k=0,\ldots,N_i-1
\end{equation}
\begin{equation} \label{Eq35g}
\hat{x}(t+k|t)\in \mathcal{T}_i^{\sigma(0|t)} \hspace{0.6cm}  for \hspace{0.2cm} k\geq \delta_i^{\sigma(0|t)}(0|t)
\end{equation}
\end{subequations}
Note that the effect of the switch is reflected in all constraints of the abovementioned optimal control problem. Constraint (\ref{Eq35g}) is a combination of constraint (\ref{Eq19f}) for the optimal control problem of DMPC and constraint (\ref{Eq27d}) for the optimal control problem of SwMPC. It ensures that before a switch occurs, the states of each subsystem in the current topology will reach a feasible region for the next topology. To guarantee stability and also feasibility when a switch occurs, a modified switch-RCI set of   for the current topology in (\ref{Eq35g}), $\mathcal{T}_i^{\sigma(0|t)}$  , must be computed as it satisfies the required conditions of a terminal region and also a switch-RCI set. This is performed by revising stage 2 of algorithm proposed in Fig.~\ref{Fig1}, i.e. by considering $\Omega_i^0\hat{\mathcal{X}}_{f_i}$ , where $\hat{\mathcal{X}}_{f_i}$  is a terminal region of the nominal subsystem (\ref{Eq32}) computed as presented in \cite{Farina19}, \cite{BettiBook31} and \cite{BettiPaper32}.
It should be noticed that when a switch occurs, all matrices include stage and terminal matrices in (\ref{Eq35}), i.e. $Q_i^0$ , $R_i^0$  and $P_i^0$ and all sets in (\ref{Eq35}) include $\mathcal{Z}_i$ , $\Delta \mathcal{E}_i$  and $\Delta \mathcal{E}_i^u$ , and the gain matrices $K_i$  of the DSwMPC have to be updated based on the new topology. Note that for the case where the next network topology is unknown in prior, $\mathcal{Z}_i^{\sigma(t)}$  is updated based on the maximum applicable disturbance. This value can be obtained with the participation of all neighbors in the allowable neighborhood set of each subsystem.

As proposed in \cite{Farina19}, \cite{BettiBook31} and \cite{BettiPaper32}, their DMPC guarantees asymptotic closed-loop stability of the origin of the distributed large-scale system and constraints satisfaction by employing the RMPC controllers as local regulators. The current research for control of distributed switched large-scale systems in which a switching signal changes the network topology proposes the DSwMPC in which the RSwMPC is employed as local regulator. The RSwMPC controllers use an implicit MPC law (\ref{Eq16}) as control input, and employ a SwMPC for a dynamic equation of nominal subsystems. As proposed in \cite{Danielson37}, the SwMPC by defining the switch-RCI sets, determines necessary and sufficient conditions for the existence of a control-law that guarantees constraint satisfaction in the presence of unknown mode switching. To ensure stability, the switch-RCI sets of SwMPC are modified as they satisfy the required conditions of a terminal region and also a switch-RCI set. In an overall view, the DSwMPC does not impose any new constraints than the original DMPC. The DSwMPC modifies the DMPC to be adopted to switch in network topology by replacing the MPC with a SwMPC controller. The SwMPC is similar to MPC only with a change in determining a terminal region. The switch-RCI set in SwMPC by modification of  $\Omega_i^0\hat{\mathcal{X}}_{f_i}$ , on one side, it has all necessary conditions of a terminal region, and the other side it can ensure a feasible switching among different modes of system. So, all discussions and results about convergence, stability and constraints satisfaction in the DMPC and the SwMPC can be generalized for the DSwMPC. In other words, by solving (\ref{Eq35}), the asymptotic stability of the closed-loop distributed switched large-scale system and constraints satisfaction are guaranteed.
\section{Results and discussions} \label{Sect6:Simulation}
In this section, the proposed DSwMPC is tested by being applied on a discrete-time distributed switched large-scale system, $\mathcal{S}$, consisting of four interacting subsystems. The models of these four subsystems are given by
\begin{eqnarray} \label{Eq36}
& \mathcal{S}_1: {x}_1(t+1)=  \begin{pmatrix}   1 & 1 \\ 0 & 1 \end{pmatrix} {x}_1(t)+
\begin{pmatrix} 0.5 \\ 1 \end{pmatrix} {u}_1(t)
+\sum_{j\in \mathcal N_1^{\sigma(t)}} A_{1j} {x}_j(t)  \nonumber  \\
& A_{12}=0.08\begin{pmatrix} 1 & 0 \\ 0 & 1 \end{pmatrix}
A_{13}=0.05\begin{pmatrix} 1 & 0 \\ 0 & 1 \end{pmatrix}
A_{14}=0.06\begin{pmatrix} 1 & 0 \\ 0 & 1 \end{pmatrix}  \nonumber \\
& \begin{pmatrix} -1.2 \\ -1.2 \end{pmatrix}  \leq \begin{pmatrix} x_{11}(t)  \\  x_{12}(t) \end{pmatrix} \leq \begin{pmatrix} 1 \\ 1 \end{pmatrix} \hspace{0.05cm} -0.5 \leq u_1(t)\leq 0.5 \nonumber
\end{eqnarray}
\begin{eqnarray} \label{Eq37}
& \mathcal{S}_2: {x}_2(t+1)=  \begin{pmatrix}   2 & -0.96 \\ 1 & 0 \end{pmatrix} {x}_2(t)+
\begin{pmatrix} 1 \\ 0 \end{pmatrix} {u}_2(t)
+\sum_{j\in \mathcal N_2^{\sigma(t)}} A_{2j} {x}_j(t)  \nonumber  \\
& A_{21}=0.05\begin{pmatrix} 1 & 0 \\ 0 & 1 \end{pmatrix}
A_{23}=0.04\begin{pmatrix} 1 & 0 \\ 0 & 1 \end{pmatrix}
A_{24}=0.04\begin{pmatrix} 1 & 0 \\ 0 & 1 \end{pmatrix}  \nonumber \\
& \begin{pmatrix} -1 \\ -1 \end{pmatrix}  \leq \begin{pmatrix} x_{21}(t)  \\  x_{22}(t) \end{pmatrix} \leq \begin{pmatrix} 1 \\ 1 \end{pmatrix} \hspace{0.05cm} -0.6 \leq u_2(t)\leq 0.6 \nonumber
\end{eqnarray}
\begin{eqnarray} \label{Eq38}
& \mathcal{S}_3: {x}_3(t+1)=  \begin{pmatrix}   1.2 & 0.51 \\ 0.1 & 1 \end{pmatrix} {x}_3(t)+
\begin{pmatrix} 0.5 \\ 1 \end{pmatrix} {u}_3(t)
+\sum_{j\in \mathcal N_3^{\sigma(t)}} A_{3j} {x}_j(t)  \nonumber  \\
& A_{31}=0.04\begin{pmatrix} 1 & 0 \\ 0 & 1 \end{pmatrix}
A_{32}=0.04\begin{pmatrix} 1 & 0 \\ 0 & 1 \end{pmatrix}
A_{34}=0.1\begin{pmatrix} 1 & 0 \\ 0 & 1 \end{pmatrix}  \nonumber \\
& \begin{pmatrix} -1.3 \\ -1.3 \end{pmatrix}  \leq \begin{pmatrix} x_{31}(t)  \\  x_{32}(t) \end{pmatrix} \leq \begin{pmatrix} 1.3 \\ 1.3 \end{pmatrix} \hspace{0.05cm} -1.1 \leq u_3(t)\leq 1.1 \nonumber
\end{eqnarray}
\begin{eqnarray} \label{Eq39}
& \mathcal{S}_4: {x}_4(t+1)=  \begin{pmatrix}   1.1 & 2 \\ 0 & 0.95 \end{pmatrix} {x}_4(t)+
\begin{pmatrix} 0 \\ 0.7787 \end{pmatrix} {u}_4(t)
+\sum_{j\in \mathcal N_4^{\sigma(t)}} A_{4j} {x}_j(t)  \nonumber  \\
& A_{41}=0.03\begin{pmatrix} 1 & 0 \\ 0 & 1 \end{pmatrix}
A_{42}=0.03\begin{pmatrix} 1 & 0 \\ 0 & 1 \end{pmatrix}
A_{43}=0.05\begin{pmatrix} 1 & 0 \\ 0 & 1 \end{pmatrix}  \nonumber \\
& \begin{pmatrix} -1.4 \\ -1.4 \end{pmatrix}  \leq \begin{pmatrix} x_{41}(t)  \\  x_{42}(t) \end{pmatrix} \leq \begin{pmatrix} 1.4 \\ 1.4 \end{pmatrix} \hspace{0.05cm} -1 \leq u_4(t)\leq 1 \nonumber
\end{eqnarray}
In the aforementioned system, after switching times, it is required that $N_i^{\sigma(t)}$ is updated based on the new network topology. Simulation time is considered equals to 15 time instances and it is assumed that the switching signal changes $N_i^{\sigma(t)}$  in $t=4$  and $t=9$ . The initial conditions are chosen to be
\begin{eqnarray*}
x_1(0)= \begin{pmatrix} -0.55 \\   0.9 \end{pmatrix} ,
x_2(0)= \begin{pmatrix} -0.8  \\  -0.9 \end{pmatrix} ,
x_3(0)= \begin{pmatrix} -0.8  \\  -0.6 \end{pmatrix} ,
x_4(0)= \begin{pmatrix} -0.7  \\   0.8 \end{pmatrix}
\end{eqnarray*}
The constrained finite-time optimal control problem of (\ref{Eq35}) is solved for each subsystem using the fmincon function in the MATLAB environment. The chosen control horizon $N_i$  and minimum dwell-time $d_i$  are 5 and 3 samples, respectively \cite{Ahandani1}. It can be noticed that the dimension of overall system does not impose any restrictions on application of the DSwMPC. But it is natural that a problem with a large dimension leads to an increase in computational time.
Our simulation study is divided into two parts. In the first part, in order to test of the DSwMPC under different work conditions and to provide the comprehensive simulations, three examples are designed based on system (\ref{Eq36})-(\ref{Eq39}) under various scenarios and the performance of DSwMPC is tested on them. In the second part, the effectiveness of DSwMPC is verified by comparing its results with those obtained by a centralized SwMPC (CSwMPC) and a DeSwMPC \cite{Ahandani1}.
\subsection{Illustrative examples} \label{SubSect6.2:examples}
\textbf{Example 1.}
Consider a distributed switched system as illustrated in (\ref{Eq36})-(\ref{Eq39}). In this scenario, let the switching times are unknown in prior but the next neighborhood sets, $N_i^{\sigma(t)}$ , are known in prior. The neighborhood set of each subsystem in three considered topologies are shown in Table~\ref{Table1} and Fig.~2. It can be noticed that it is not necessary to have the same number of neighbors for all the subsystems and size of  $N_i^{\sigma(t)}$ in each subsystem can be different. The size of  $N_i^{\sigma(t)}$ can be restricted by (\ref{Eq33}) and (\ref{Eq34}) where a big neighborhood set imposes a big $\mathcal{Z}_i^{\sigma(t)}$  and so it is possible that the tightened constraints of $\hat{\mathcal{X}}_i^{\sigma(t)}$  and $\hat{\mathcal{U}}_i^{\sigma(t)}$  be the empty sets.
\begin{table}
\caption{Neighborhood set of each subsystem in three considered topologies.}
\label{Table1}
\begin{center}
\begin{tabular}{c c c c c}
  \hline
  Time interval & $N_{1}$ & $N_{2}$ & $N_{3}$ & $N_{4}$ \\
  0-4 & \{2\} & \{1\} & \{4\} & \{3\} \\
  5-9 & \{2,3\} & \{3,4\} & \{3,4\} & \{1,3\} \\
  10-14 & \{4\} & \{4\} & \{2\} & \{1\} \\
  \hline
\end{tabular}
\end{center}
\end{table}
\begin{figure*}
\begin{center}
\begin{small}
\includegraphics[scale=0.7]{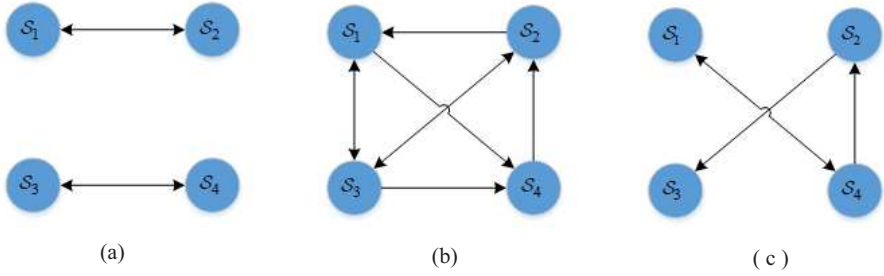}    
\caption{Three considered network topologies. (a) Initial topology, (b) Second topology, (c) Third topology}  
\end{small}
\end{center}
\label{Fig2}
\end{figure*}
The state responses and the inputs of the four subsystems are presented in Fig.~\ref{Fig3} and Fig.~\ref{Fig4}. It can be observed that constraints on states and inputs are always fulfilled by the proposed DSwMPC. The simulation results also confirm that the closed-loop system successfully converges to the origin.

\textbf{Example 2.}
In Example 2, let both switching times and the next neighboring sets to be unknown in prior. All simulation parameters and characteristics are the same with Example 1. Only the possible neighbors of different subsystems are restricted to be:
$N_1^{\sigma(t)}=\{2,3\}$ , $N_2^{\sigma(t)}=\{1,3,4\}$ , $N_3^{\sigma(t)}=\{1,2\}$  and $N_4^{\sigma(t)}=\{1,3\}$  where for example $N_1^{\sigma(t)}=\{2,3\}$  means that the neighbors of subsystem 1 under the switching signal   can be subsystem 2 or 3 or both of them. Since the neighborhood sets have been supposed to be unknown in prior, the maximum RPI set $\mathcal{Z}_i$ , $i=1,\ldots,4$ , of each subsystem must be computed corresponding the maximum applicable disturbance to it in all possible topologies. For example for $\mathcal{S}_1$ , the maximum RPI set $\mathcal{Z}_1$  is computed with this assumption that its neighborhood set is composed of both subsystems 2 and 3. It can be computed off-line and is used to provide a feasible switch among topologies.

\begin{figure*}[ht]
  \subfloat[subsystem 1]{
	\begin{subfigure}[c][1\width]{0.4\textwidth}
	   \centering
	   \includegraphics[width=1\textwidth]{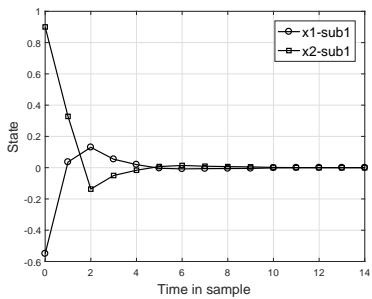}
	\end{subfigure}}
 \hfill 	
  \subfloat[subsystem 2]{
	\begin{subfigure}[c][1\width]{0.4\textwidth}
	   \centering
	   \includegraphics[width=1\textwidth]{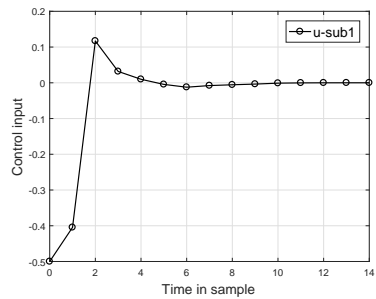}
	\end{subfigure}}
 \hfill	
  \subfloat[subsystem 3]{
	\begin{subfigure}[c][1\width]{0.4\textwidth}
	   \centering
	   \includegraphics[width=1\textwidth]{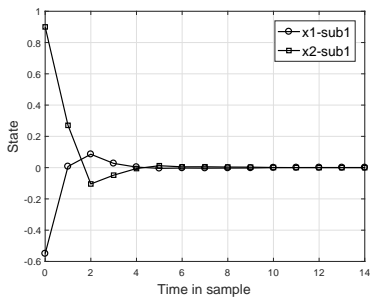}
	\end{subfigure}}
 \hfill	
  \subfloat[subsystem 4]{
	\begin{subfigure}[c][1\width]{0.4\textwidth}
	   \centering
	   \includegraphics[width=1\textwidth]{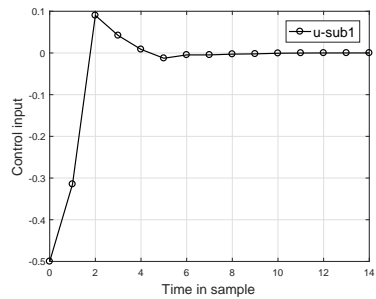}
	\end{subfigure}}
\caption{The response curves of the states trajectories in Example 1.}
\label{Fig3}
\end{figure*}

\begin{figure*}[ht]
  \subfloat[subsystem 1]{
	\begin{subfigure}[c][1\width]{0.4\textwidth}
	   \centering
	   \includegraphics[width=1\textwidth]{J2_Phd_Fig4a.eps}
	\end{subfigure}}
 \hfill 	
  \subfloat[subsystem 2]{
	\begin{subfigure}[c][1\width]{0.4\textwidth}
	   \centering
	   \includegraphics[width=1\textwidth]{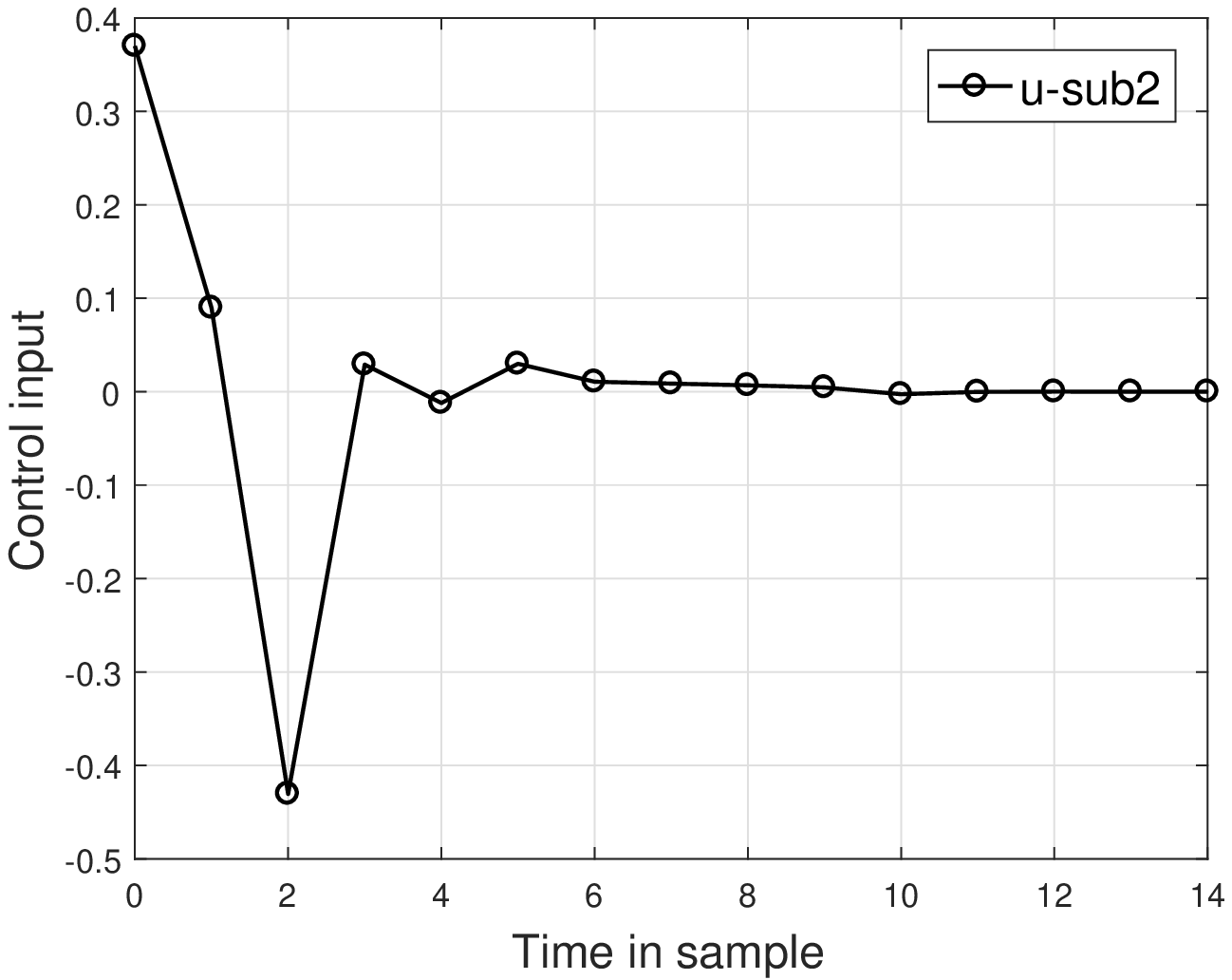}
	\end{subfigure}}
 \hfill	
  \subfloat[subsystem 3]{
	\begin{subfigure}[c][1\width]{0.4\textwidth}
	   \centering
	   \includegraphics[width=1\textwidth]{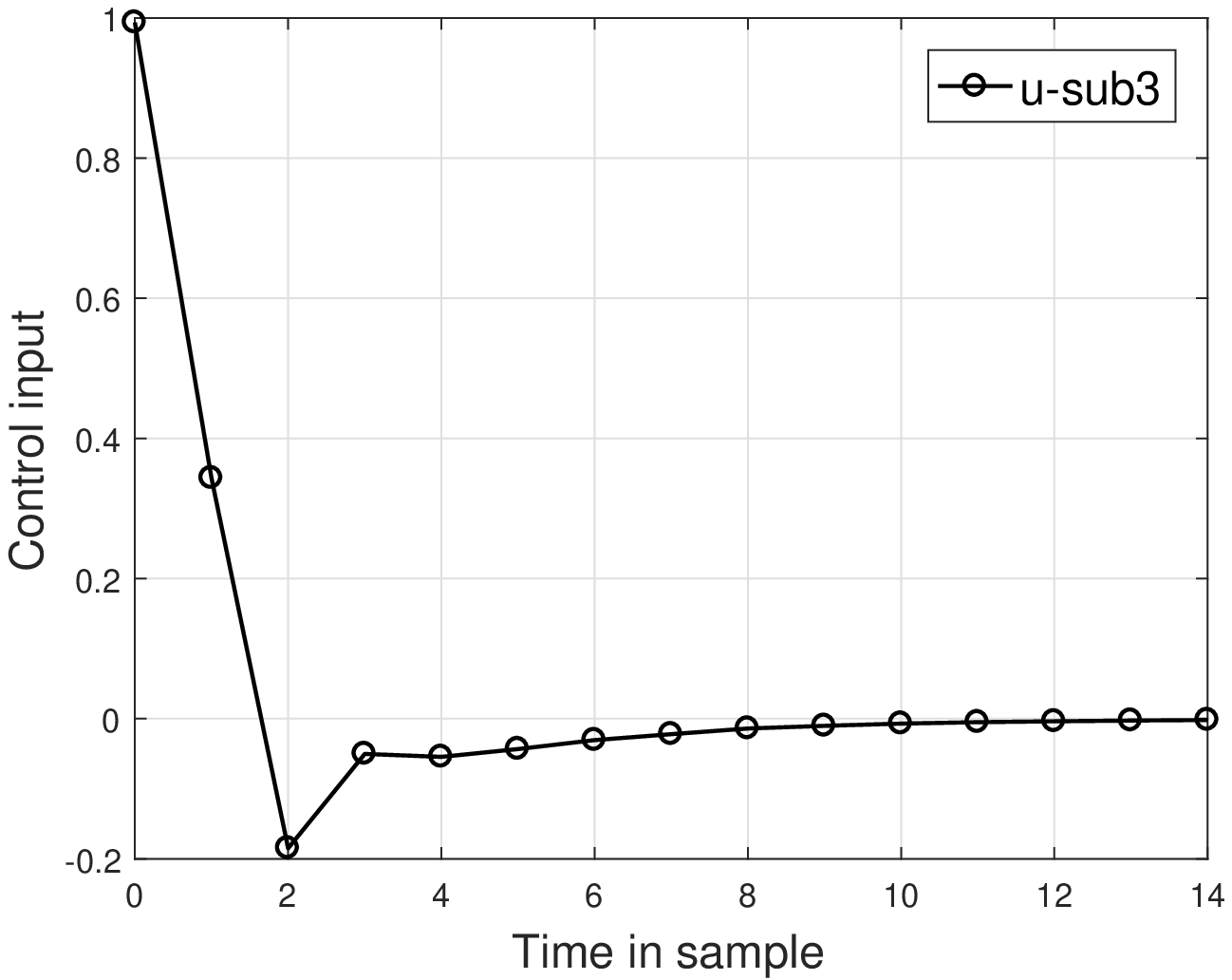}
	\end{subfigure}}
 \hfill	
  \subfloat[subsystem 4]{
	\begin{subfigure}[c][1\width]{0.4\textwidth}
	   \centering
	   \includegraphics[width=1\textwidth]{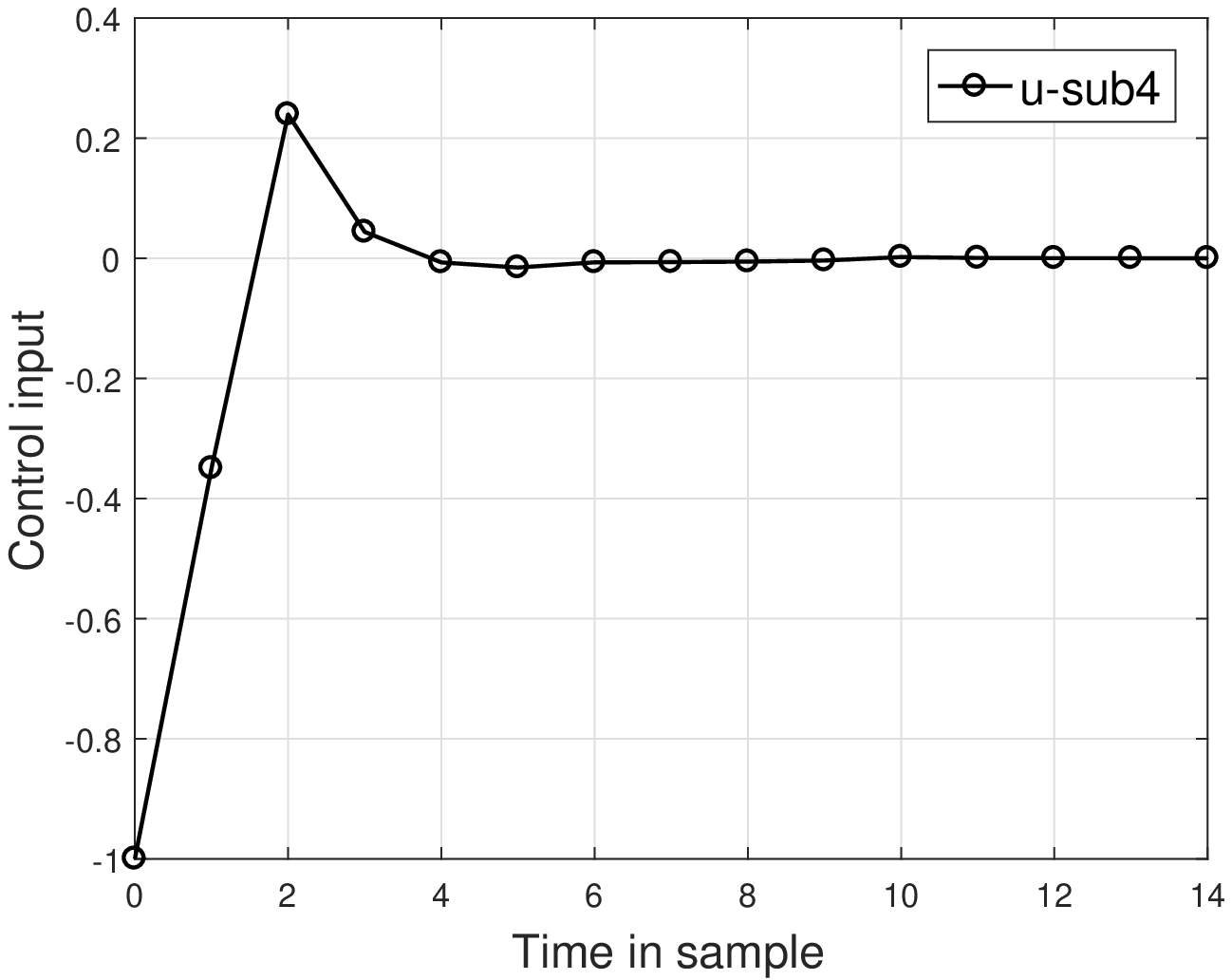}
	\end{subfigure}}
\caption{The response curves of the input trajectories in Example 1.}
\label{Fig4}
\end{figure*}

\begin{figure*}[ht]
  \subfloat[subsystem 1]{
	\begin{subfigure}[c][1\width]{0.4\textwidth}
	   \centering
	   \includegraphics[width=1\textwidth]{J2_Phd_Fig5a.eps}
	\end{subfigure}}
 \hfill 	
  \subfloat[subsystem 2]{
	\begin{subfigure}[c][1\width]{0.4\textwidth}
	   \centering
	   \includegraphics[width=1\textwidth]{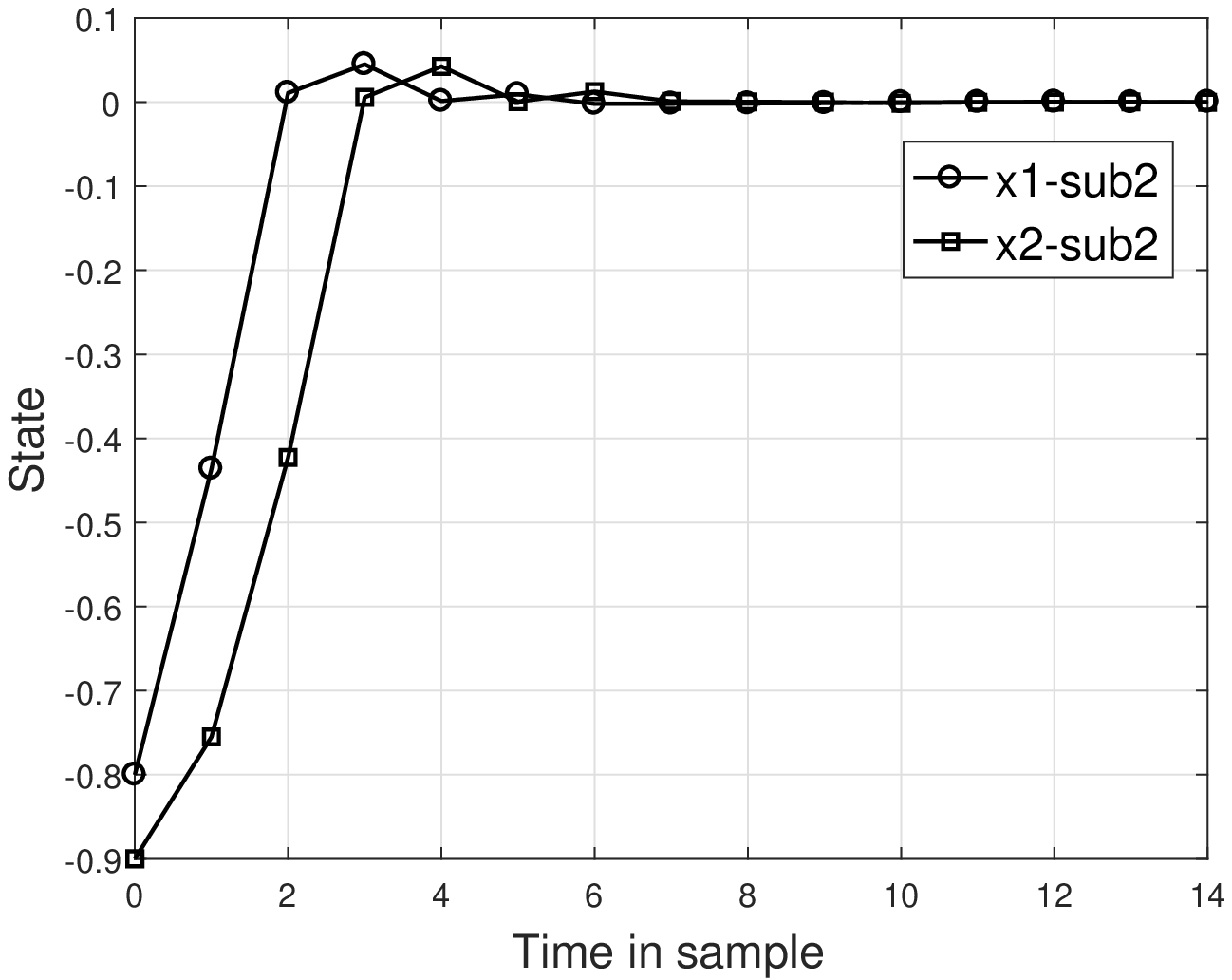}
	\end{subfigure}}
 \hfill	
  \subfloat[subsystem 3]{
	\begin{subfigure}[c][1\width]{0.4\textwidth}
	   \centering
	   \includegraphics[width=1\textwidth]{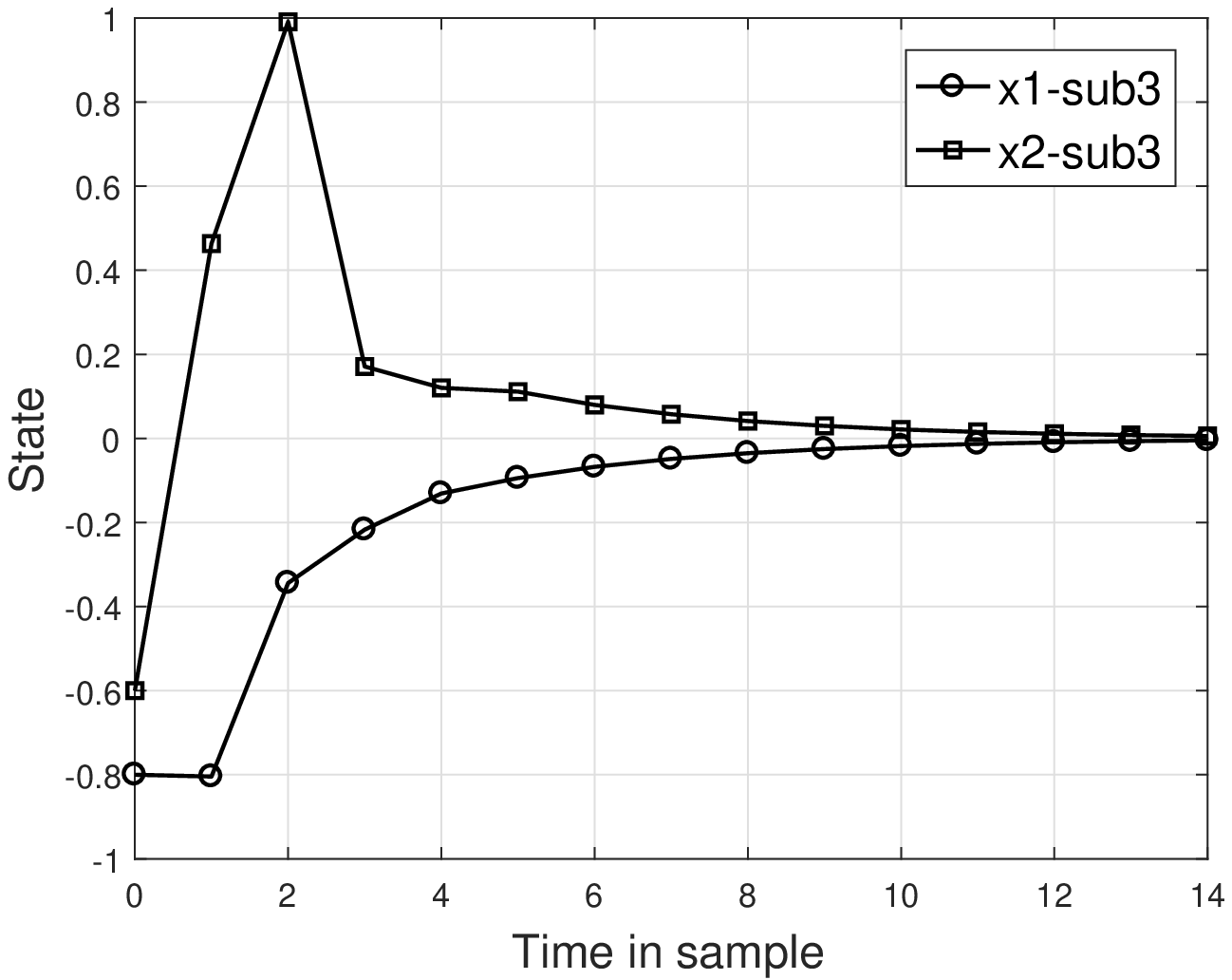}
	\end{subfigure}}
 \hfill	
  \subfloat[subsystem 4]{
	\begin{subfigure}[c][1\width]{0.4\textwidth}
	   \centering
	   \includegraphics[width=1\textwidth]{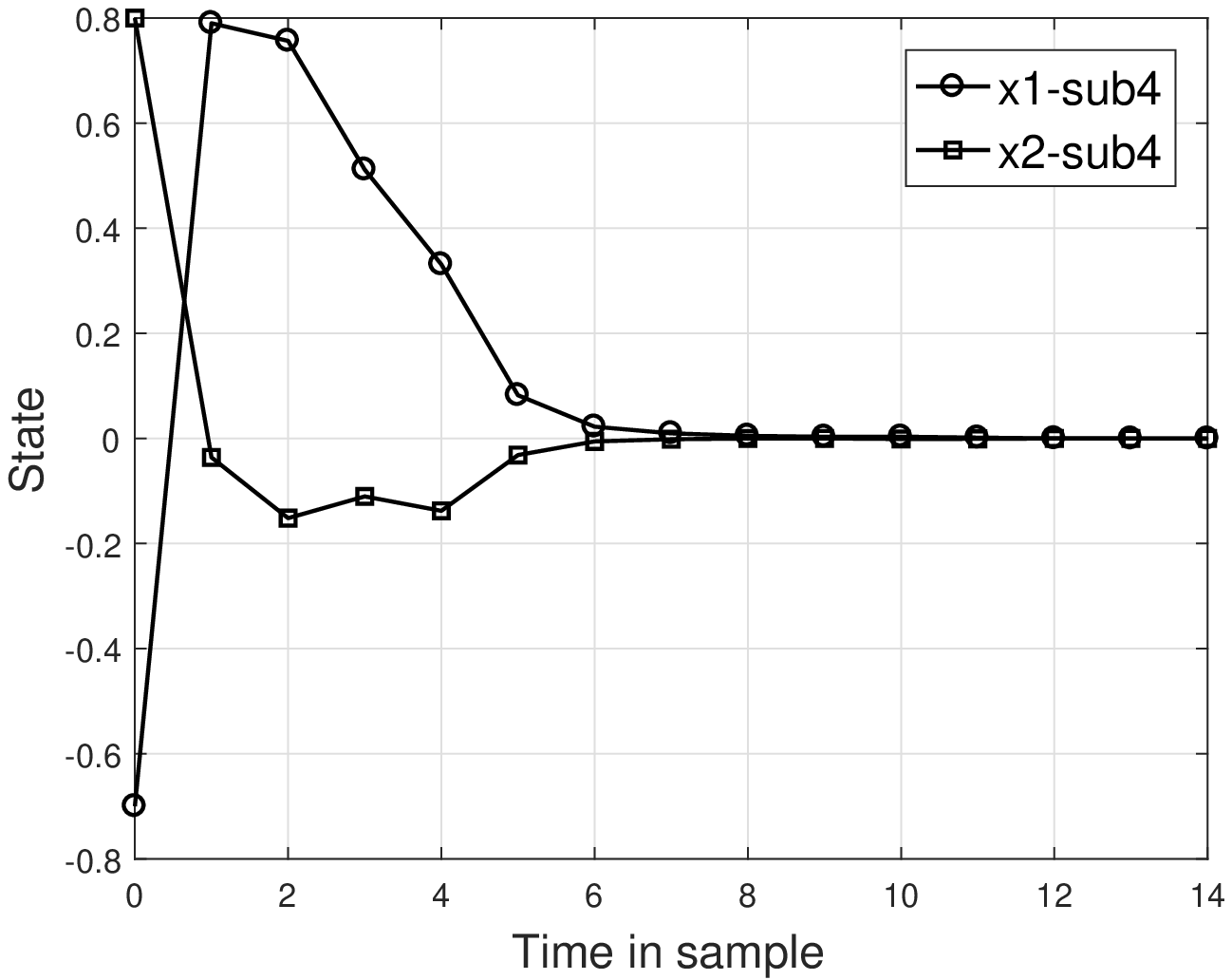}
	\end{subfigure}}
\caption{The response curves of the states trajectories in Example 2.}
\label{Fig5}
\end{figure*}

\begin{figure*}[ht]
  \subfloat[subsystem 1]{
	\begin{subfigure}[c][1\width]{0.4\textwidth}
	   \centering
	   \includegraphics[width=1\textwidth]{J2_Phd_Fig6a.eps}
	\end{subfigure}}
 \hfill 	
  \subfloat[subsystem 2]{
	\begin{subfigure}[c][1\width]{0.4\textwidth}
	   \centering
	   \includegraphics[width=1\textwidth]{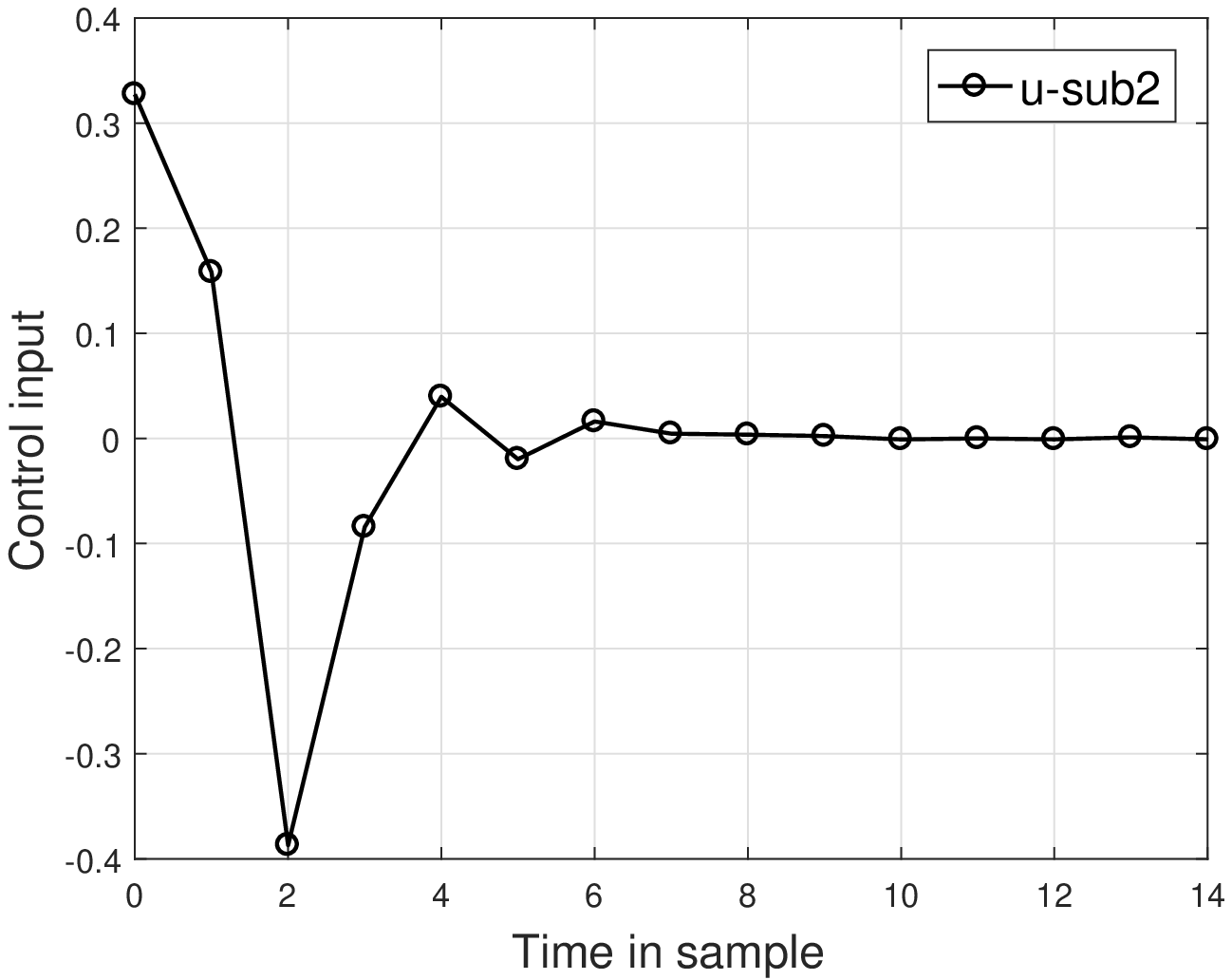}
	\end{subfigure}}
 \hfill	
  \subfloat[subsystem 3]{
	\begin{subfigure}[c][1\width]{0.4\textwidth}
	   \centering
	   \includegraphics[width=1\textwidth]{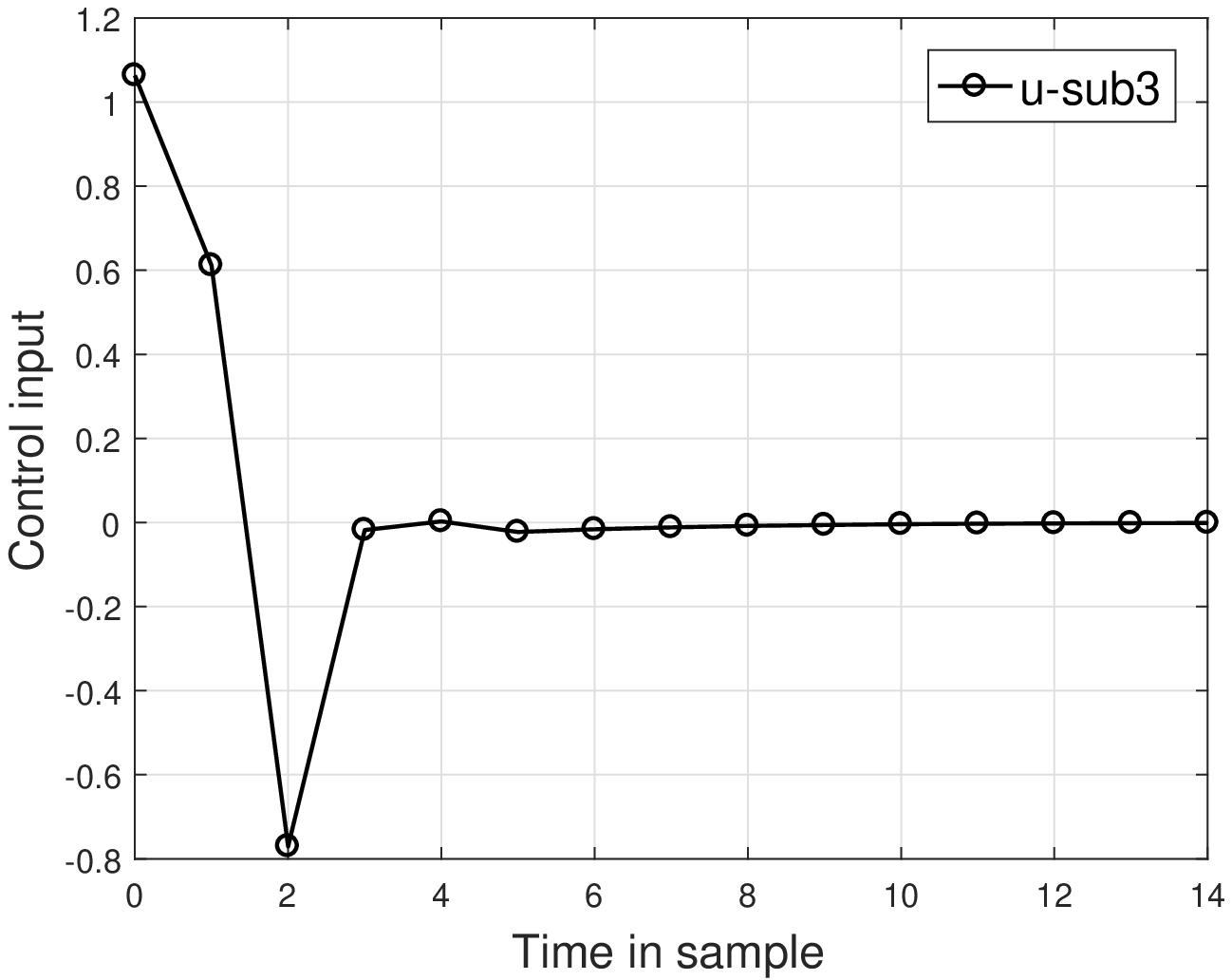}
	\end{subfigure}}
 \hfill	
  \subfloat[subsystem 4]{
	\begin{subfigure}[c][1\width]{0.4\textwidth}
	   \centering
	   \includegraphics[width=1\textwidth]{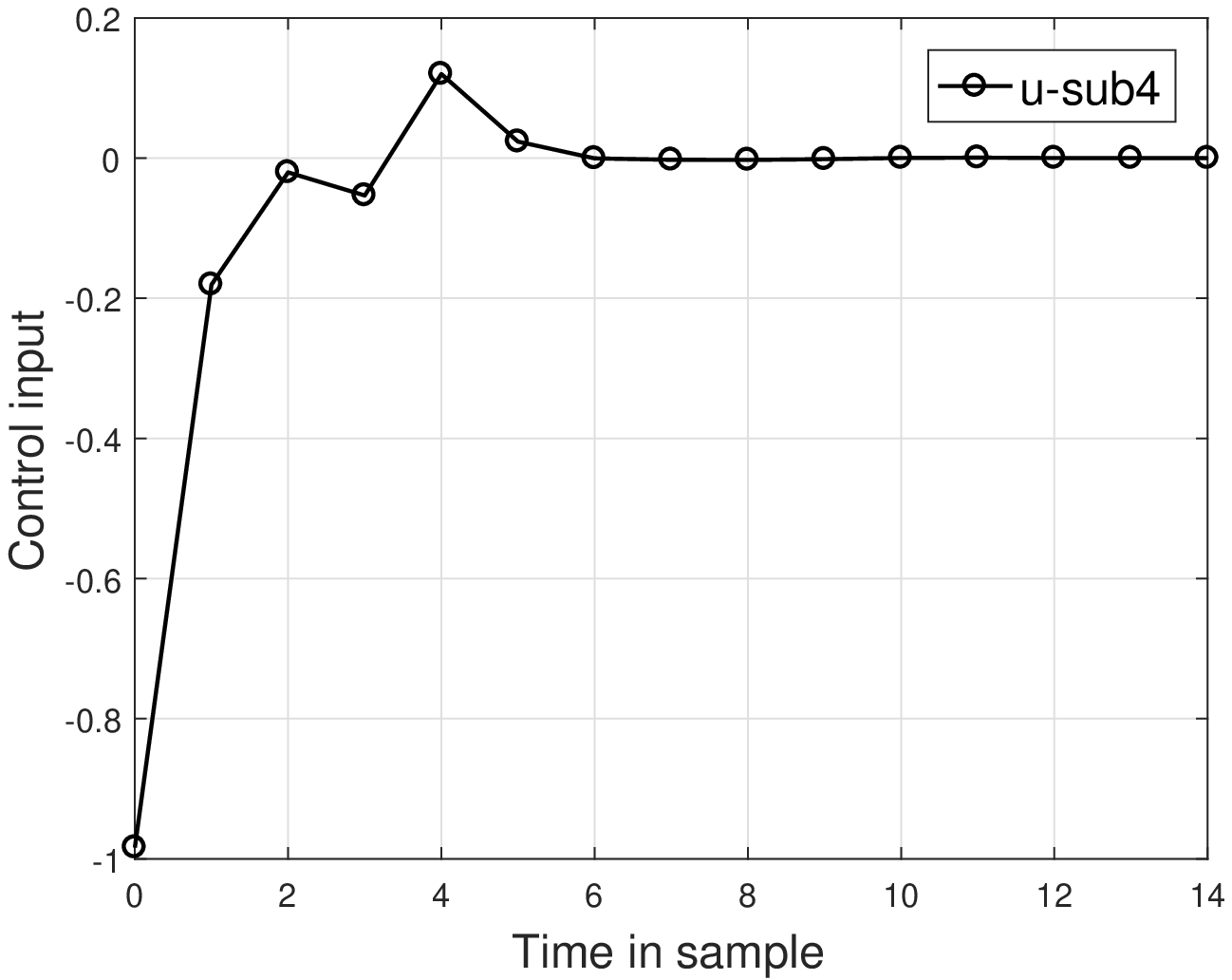}
	\end{subfigure}}
\caption{The response curves of the input trajectories in Example 2.}
\label{Fig6}
\end{figure*}

The evolutions of states and control input of the four subsystems for Example 2 are depicted in Fig.~\ref{Fig5} and Fig.~\ref{Fig6}, respectively. The achieved results show that under this control strategy the closed-loop system converges to the origin. In other words, they confirm that the DSwMPC successfully stabilizes closed-loop system. Clearly, it can be observed that the constraints on the input and states are satisfied by the DSwMPC at all times. However, in comparison with Example 1, the convergence speed of curves is slower than the corresponding curves in Fig.~\ref{Fig3} and Fig.~\ref{Fig4}. These observations as mentioned above, clearly verify the robustness characteristics of the DSwMPC against unknown switching times and especially against unknown next topology.

\begin{figure*}[ht]
  \subfloat[subsystem 1]{
	\begin{subfigure}[c][1\width]{0.4\textwidth}
	   \centering
	   \includegraphics[width=1\textwidth]{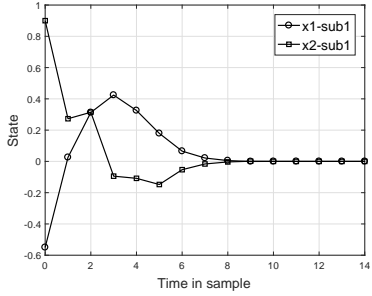}
	\end{subfigure}}
 \hfill 	
  \subfloat[subsystem 2]{
	\begin{subfigure}[c][1\width]{0.4\textwidth}
	   \centering
	   \includegraphics[width=1\textwidth]{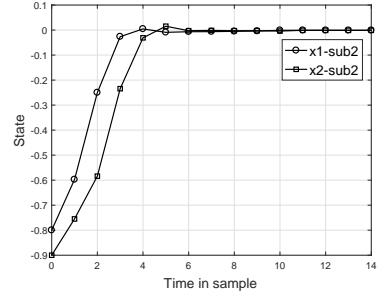}
	\end{subfigure}}
 \hfill	
  \subfloat[subsystem 3]{
	\begin{subfigure}[c][1\width]{0.4\textwidth}
	   \centering
	   \includegraphics[width=1\textwidth]{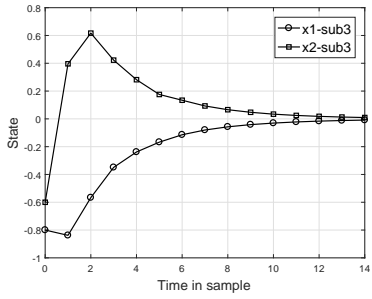}
	\end{subfigure}}
 \hfill	
  \subfloat[subsystem 4]{
	\begin{subfigure}[c][1\width]{0.4\textwidth}
	   \centering
	   \includegraphics[width=1\textwidth]{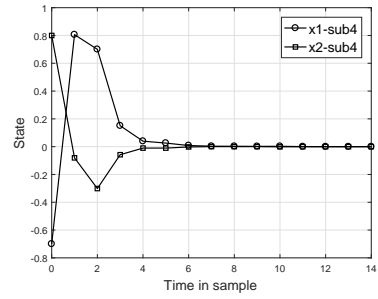}
	\end{subfigure}}
\caption{The response curves of the states trajectories in Example 3.}
\label{Fig7}
\end{figure*}

\begin{figure*}[ht]
  \subfloat[subsystem 1]{
	\begin{subfigure}[c][1\width]{0.4\textwidth}
	   \centering
	   \includegraphics[width=1\textwidth]{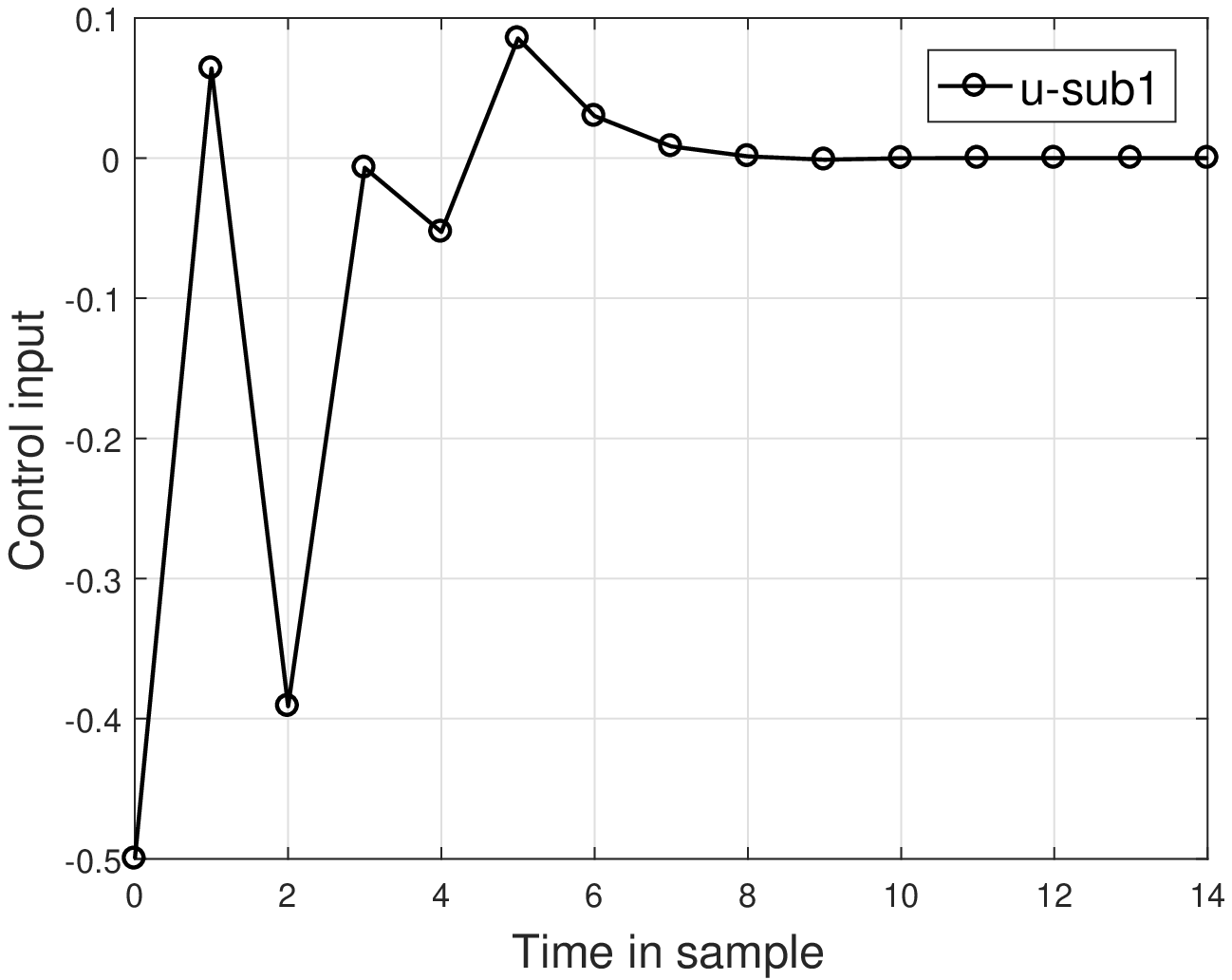}
	\end{subfigure}}
 \hfill 	
  \subfloat[subsystem 2]{
	\begin{subfigure}[c][1\width]{0.4\textwidth}
	   \centering
	   \includegraphics[width=1\textwidth]{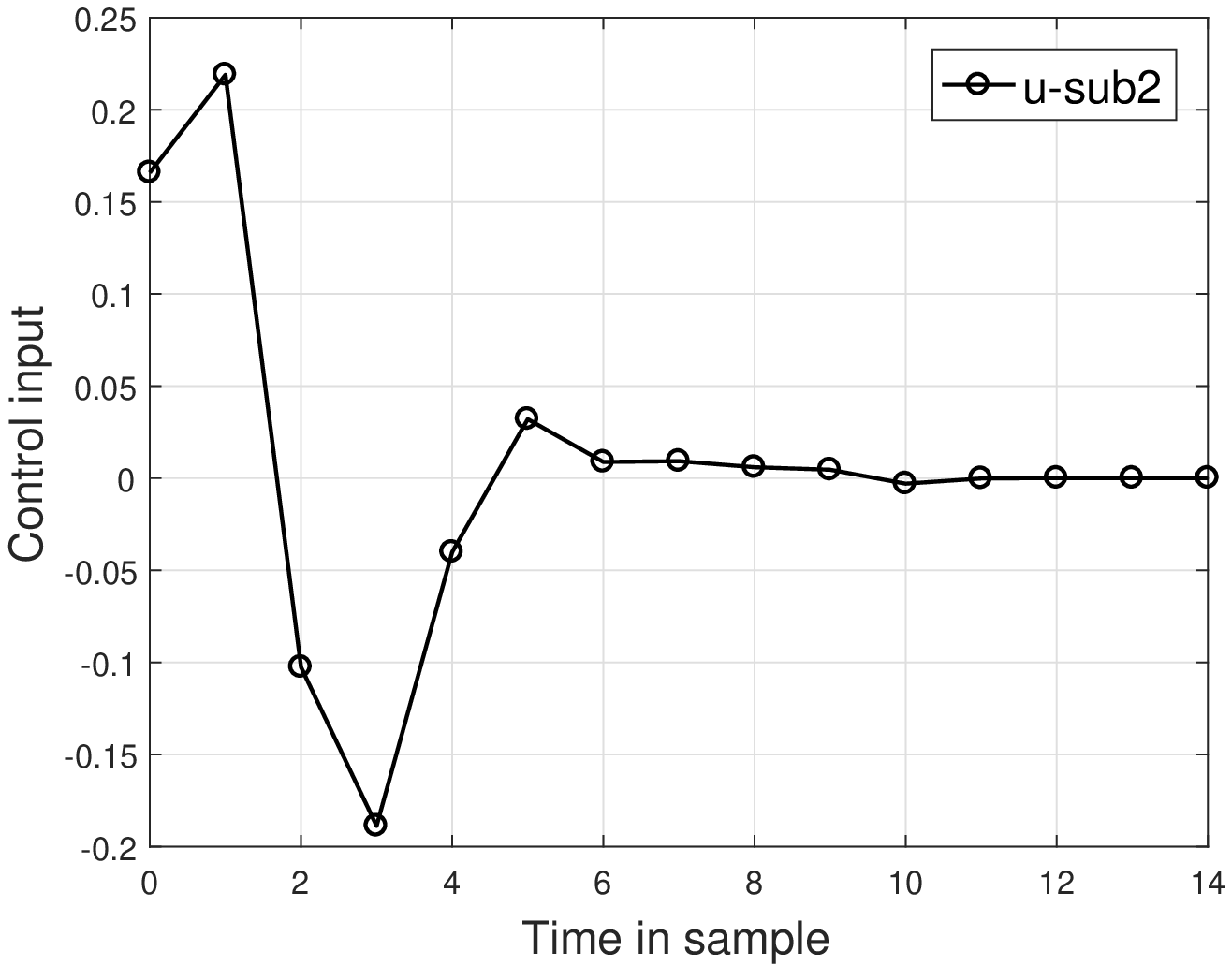}
	\end{subfigure}}
 \hfill	
  \subfloat[subsystem 3]{
	\begin{subfigure}[c][1\width]{0.4\textwidth}
	   \centering
	   \includegraphics[width=1\textwidth]{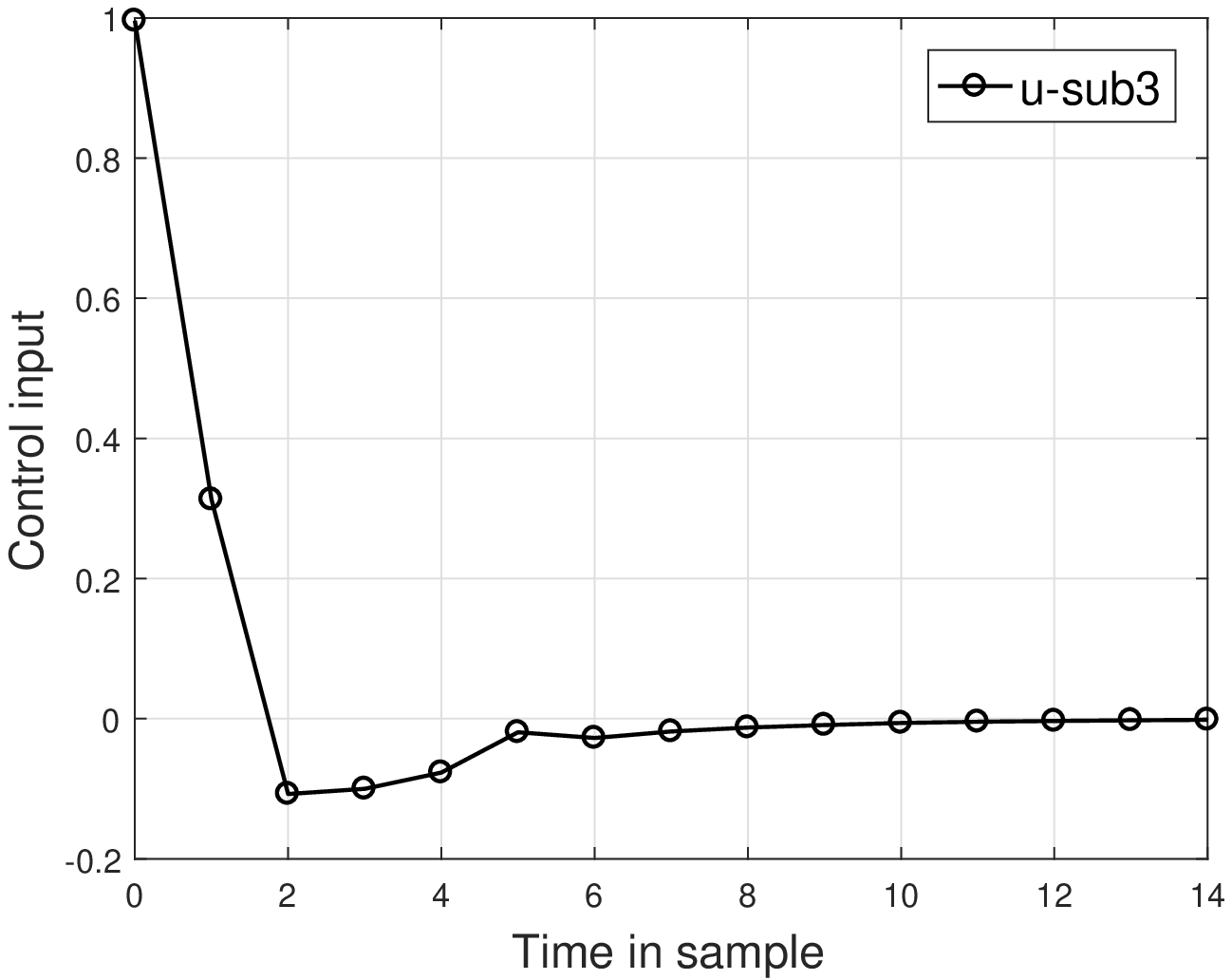}
	\end{subfigure}}
 \hfill	
  \subfloat[subsystem 4]{
	\begin{subfigure}[c][1\width]{0.4\textwidth}
	   \centering
	   \includegraphics[width=1\textwidth]{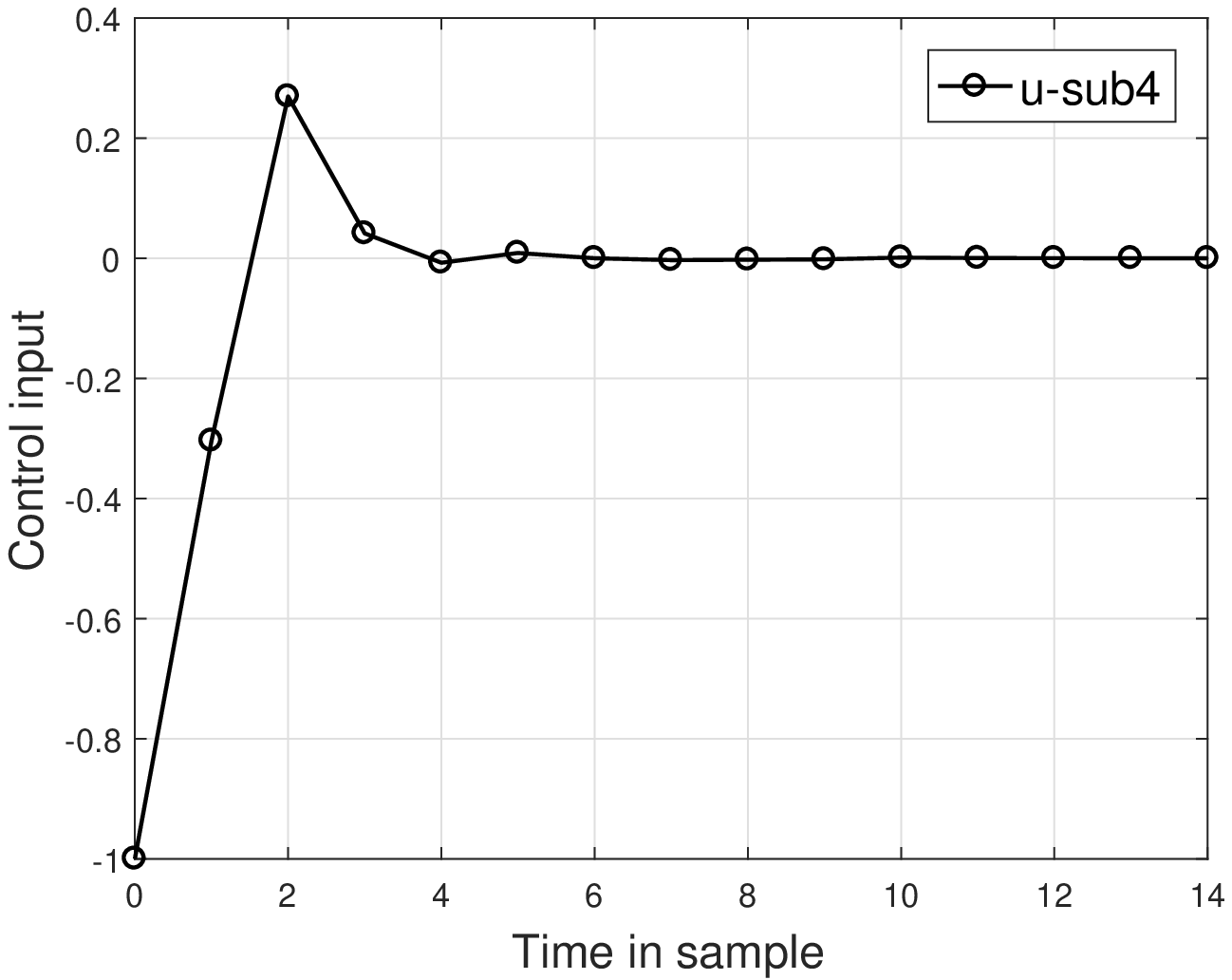}
	\end{subfigure}}
\caption{The response curves of the input trajectories in Example 3.}
\label{Fig8}
\end{figure*}

\textbf{Example 3.}
In this example such Example 2, it is assumed that both of switching time and the neighborhood sets, $N_i^{\sigma(t)}$ , after switch are unknown in prior. The neighborhood set of each subsystem in every topology and switching times are the same as Example 1. In comparison with Example 2, the sets of possible neighborhoods are  unknown in this example and the controller have no information about the next possible topologies. In other words, it is assumed that $N_1^{\sigma(t)}=\{2,3,4\}$ , $N_2^{\sigma(t)}=\{1,3,4\}$ , $N_3^{\sigma(t)}=\{1,2,4\}$ and $N_4^{\sigma(t)}=\{1,2,3\}$  where for example $N_1^{\sigma(t)}=\{2,3,4\}$  means that the neighbors of subsystem 1 under switching signal $\sigma(t)$  can be subsystem 2 or 3 or 4 or a combination of two or three neighbors. So such Example 2, the maximum RPI set $\mathcal{Z}_i$ , $i=1,\ldots,4$ , of each subsystem must be computed corresponding the maximum applicable disturbance to it in all possible topologies. It means that the maximum applicable disturbance to each subsystem can be computed by assumption that it is impacted by all of the other subsystems, e.g. for subsystem 1 we have $\mathcal{W}_1^\sigma(t)=\bigoplus_{j\in {\mathcal N}_1^{\sigma(t)}} (A_{1j}\mathcal{E}_j+B_{1j}\mathcal{E}_j^u)$  .
The simulation results for evolutions of states and control input on Example 3 are depicted in Fig.~\ref{Fig7} and Fig.~\ref{Fig8}, respectively. It can be observed that despite considering the maximum applicable disturbance on each subsystem, the constraints on the input and states are addresses by the DSwMPC at all times. Moreover, as again shown by these figures, the closed-loop system converges to the origin. However, the convergence speed of curves is less than the corresponding curves in Fig.~\ref{Fig3} to Fig.~\ref{Fig6}. These figures in comparison with Example 2, clearly show the robustness characteristics of the DSwMPC against strong interactions among subsystems. Also Table ~\ref{Table2} summarizes the numerical comparisons accrording to the state square errors of close-loop system under the control of the DSWMPC for Examples 1-3. As the obtained results of this table show, the value of the sum of square errors increases from Example 1 to Example 3.

\begin{table*}
\caption{State square errors of DSwMPC for Examples 1-3.}
\label{Table2}
\centering
\resizebox{\textwidth}{!}{%
\begin{tabular}{c p{0.01cm} c c p{0.01cm} c c p{0.01cm} c c p{0.01cm} c c p{0.01cm} c}
  \hline
   Example & & \multicolumn{2}{c}{Sub1} & & \multicolumn{2}{c}{Sub2} & & \multicolumn{2}{c}{Sub3} & & \multicolumn{2}{c}{Sub4} & & Sum \\ \cline{3-4} \cline{6-7} \cline{9-10} \cline{12-13}
            & &$x_{1}$ & $x_{2}$& &$x_{1}$& $x_{2}$ & & $x_{1}$& $x_{2}$& &$x_{1}$& $x_{2}$& & \\ \hline
  Example 1 & & 0.5694 & 0.9693 & & 0.8921 & 1.2341 & & 1.3781 & 1.1156 & & 1.2964 & 0.8580 & & 8.3129 \\
  Example 2 & & 0.5574 & 0.9470 & & 0.9124 & 1.2493 & & 1.2198 & 1.2752 & & 1.4370 & 0.8346 & & 8.4326 \\
  Example 3 & & 0.8510 & 1.0143 & & 1.0298 & 1.3331 & & 1.3771 & 1.1054 & & 1.2866 & 0.8607 & & 8.8579 \\
  \hline
\end{tabular}
}
\end{table*}

\subsection{Comparison study} \label{SubSect6.2:comparison}
To demonstrate the performance of the proposed DSwMPC more clearly, a comparison of the DSwMPC and results of DeSwMPC \cite{Ahandani1} and the CSwMPC is also provided.
The CSwMPC has been designed as proposed in Section 4. All parameter values of CSwMPC are considered equal to parameters of the local SwMPC. This comparison study is carried out using Example 1 to Example 3. To apply the CSwMPC of Section 4 on a system, it is necessary that the operating modes of system to be known in advance (see algorithm of Fig.~\ref{Fig1}). Since in Examples 2 and 3 the subsequent topology is not known in advance, so subsequent operating modes of the overall system are not known and this version of CSwMPC cannot be applied on such a system. For the DeSwMPC, all parameters were tuned based on those proposed in \cite{Ahandani1}.

The trajectories of the control inputs and the states associated with the four subsystems are depicted in Fig.~\ref{Fig9} to Fig.~\ref{Fig12}. Also, the sum of square errors of close-loop system for the DSwMPC, DeSwMPC and CSwMPC are shown in Table.~\ref{Table3}. It can be observed that the shapes of the state response curves under the CSwMPC are very similar to those obtained by the DSwMPC. Only the states of closed loop system under CSwMPC are a bit faster with respect to the DSwMPC, especially for states of subsystems 2 and 3. The achieved results of Table.~\ref{Table3} highlight the effectiveness of  DSwMPC. However, the CSwMPC has the best quality in Example 1 but in addition to overall problems of centralized control strategies, it can not be applied on systems with scenarios such as Example 2 and Example 3.  Also about the DecSwMPC, it is observed that however it achieves a faster convergence in Example 1, but its performance is dramatically decreased in Examples 2 and 3 when number of agents in the sets of possible neighborhoods is increased so that this method can not find any feasible control actions on Example 3. From these results we can see that, compared to the CSwMPC and the DeSwMPC proposed in \cite{Ahandani1}, the DSwMPC proposed in this article has a considerable flexibility to be applied on distributed switched systems in which the system topology changes under various scenarios.

\begin{figure*}[ht]
  \subfloat[subsystem 1]{
	\begin{subfigure}[c][1\width]{0.4\textwidth}
	   \centering
	   \includegraphics[width=1\textwidth]{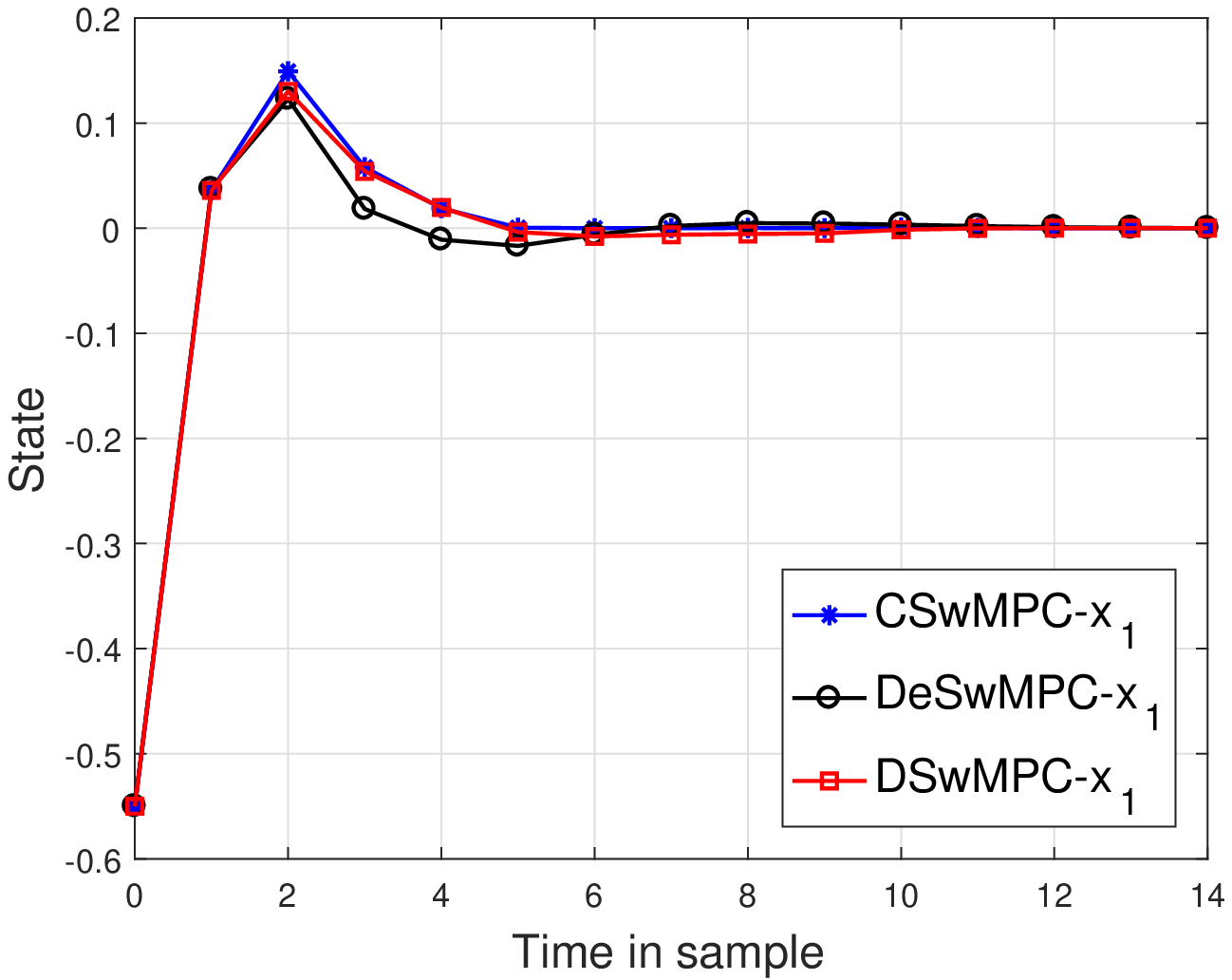}
	\end{subfigure}}
 \hfill 	
  \subfloat[subsystem 1]{
	\begin{subfigure}[c][1\width]{0.4\textwidth}
	   \centering
	   \includegraphics[width=1\textwidth]{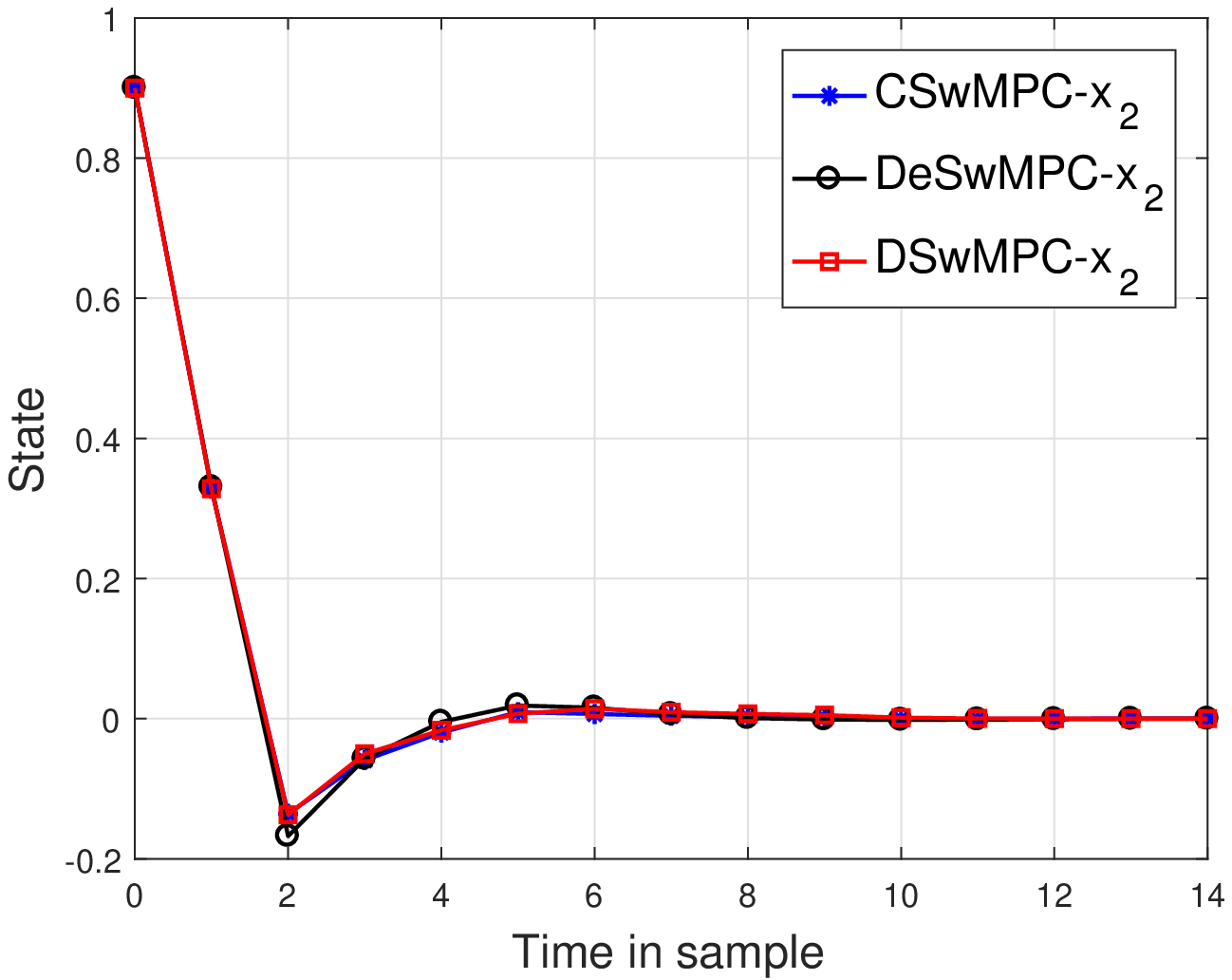}
	\end{subfigure}}
 \hfill	
  \subfloat[subsystem 2]{
	\begin{subfigure}[c][1\width]{0.4\textwidth}
	   \centering
	   \includegraphics[width=1\textwidth]{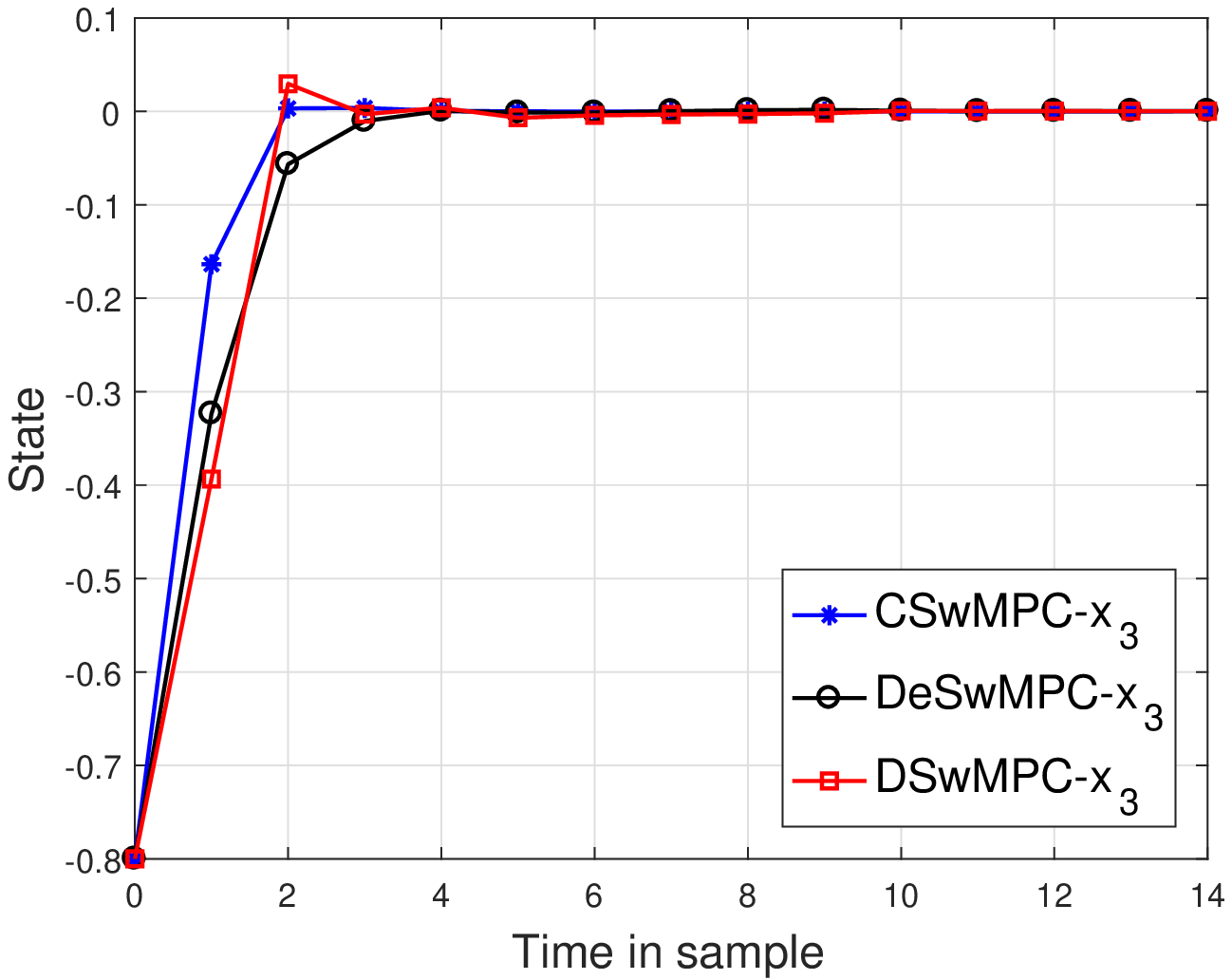}
	\end{subfigure}}
 \hfill	
  \subfloat[subsystem 2]{
	\begin{subfigure}[c][1\width]{0.4\textwidth}
	   \centering
	   \includegraphics[width=1\textwidth]{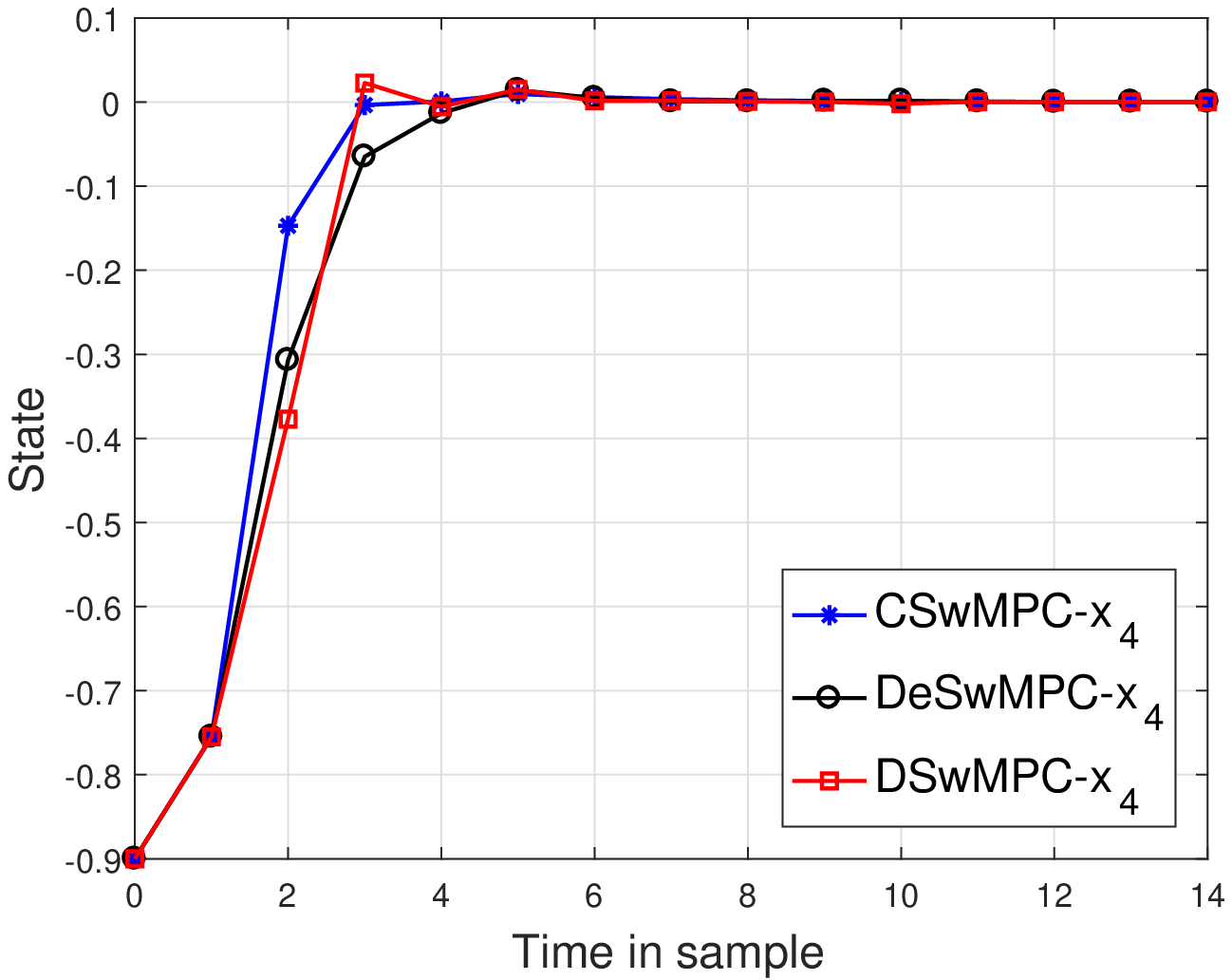}
	\end{subfigure}}
 \hfill	
  \subfloat[subsystem 3]{
	\begin{subfigure}[c][1\width]{0.4\textwidth}
	   \centering
	   \includegraphics[width=1\textwidth]{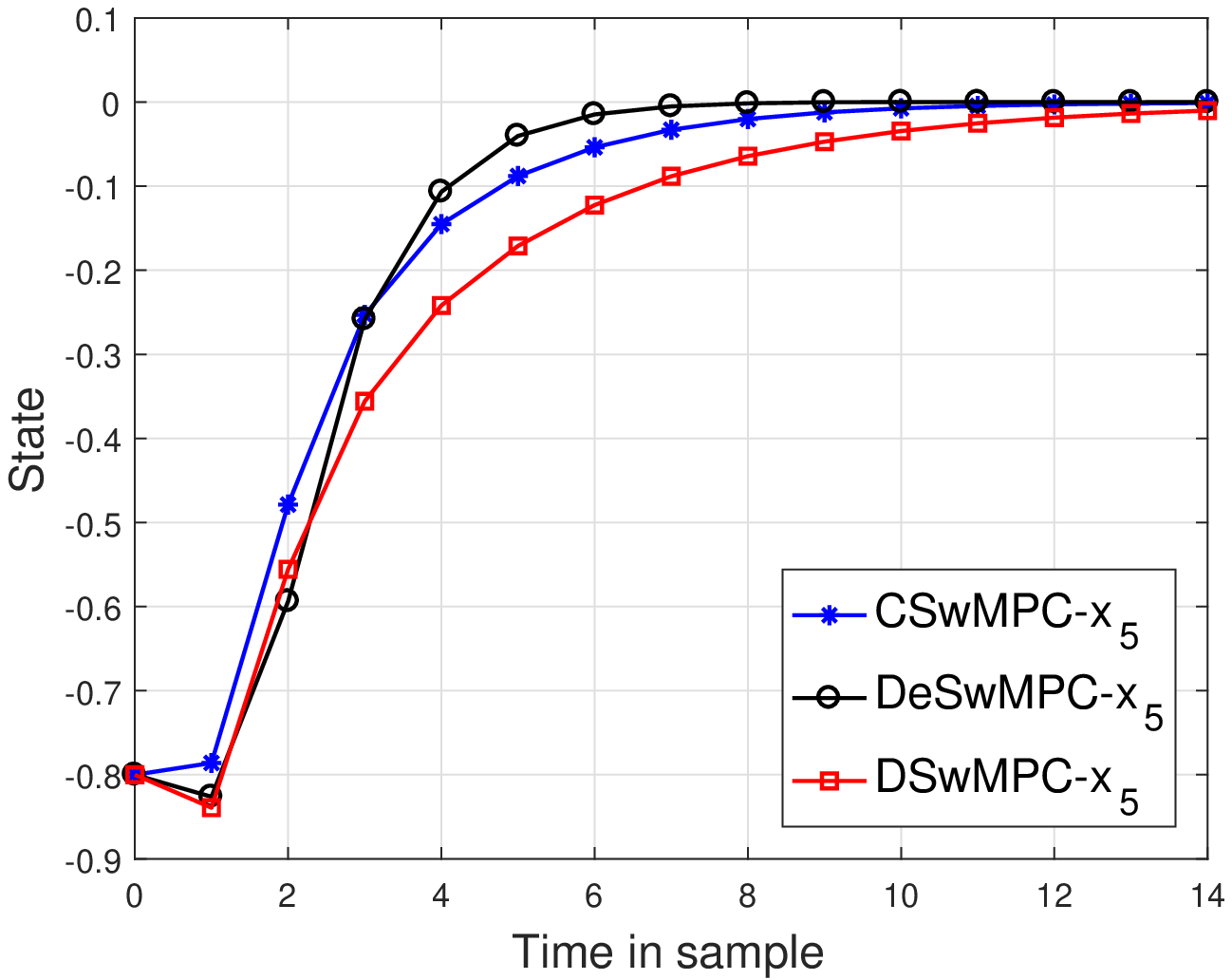}
	\end{subfigure}}
 \hfill	
  \subfloat[subsystem 3]{
	\begin{subfigure}[c][1\width]{0.4\textwidth}
	   \centering
	   \includegraphics[width=1\textwidth]{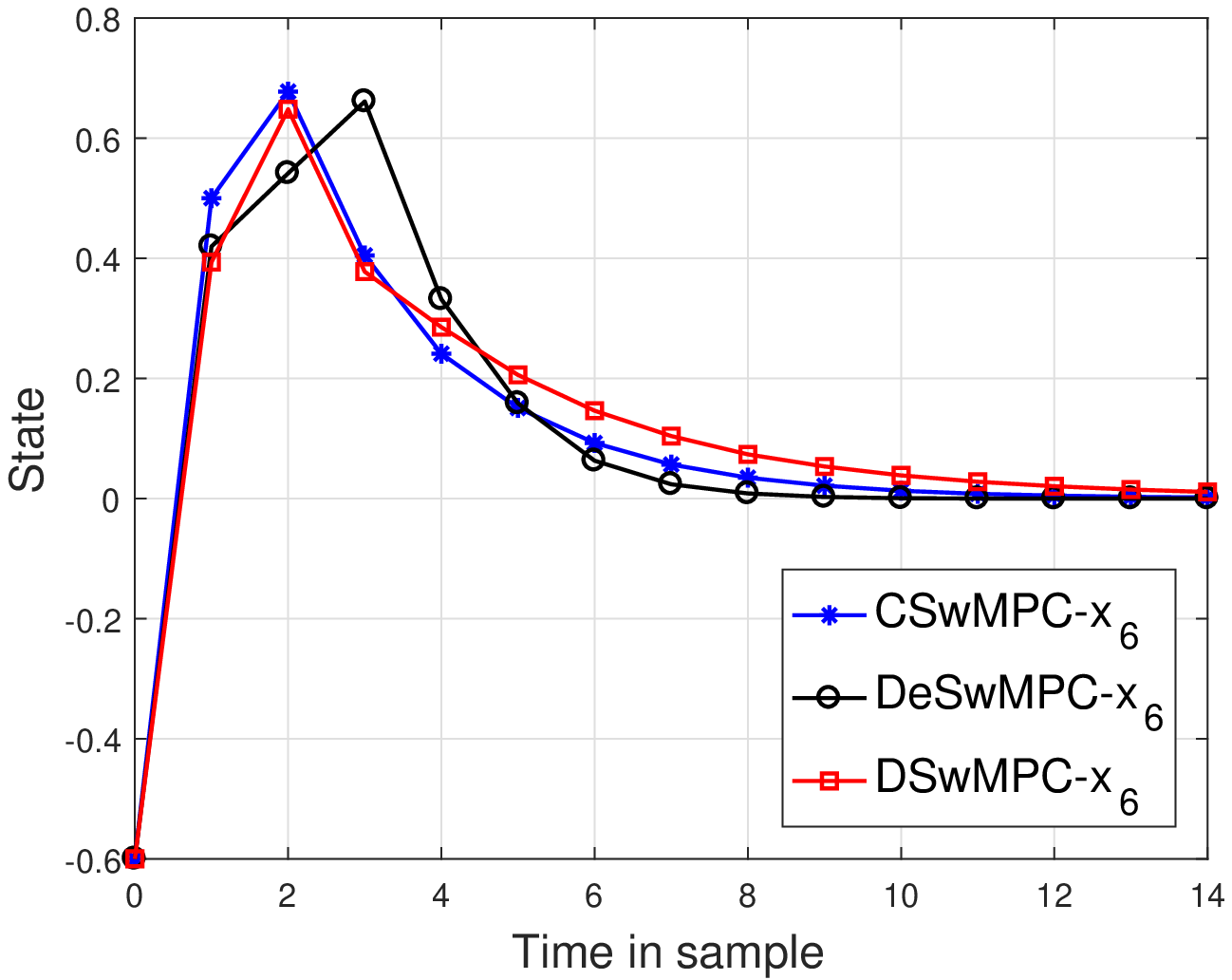}
	\end{subfigure}}
\end{figure*}
\begin{figure*}[ht]
\ContinuedFloat	
  \subfloat[subsystem 4]{
	\begin{subfigure}[c][1\width]{0.4\textwidth}
	   \centering
	   \includegraphics[width=1\textwidth]{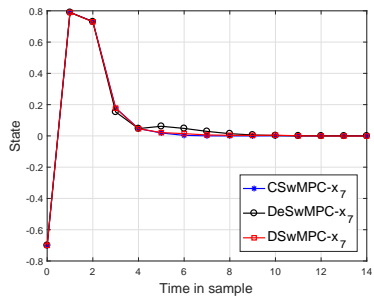}
	\end{subfigure}}
 \hfill	
  \subfloat[subsystem 4]{
	\begin{subfigure}[c][1\width]{0.4\textwidth}
	   \centering
	   \includegraphics[width=1\textwidth]{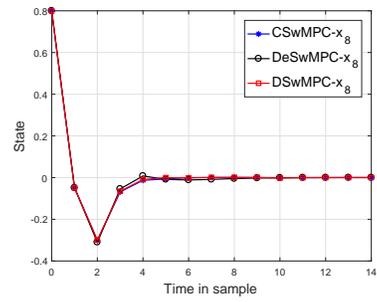}
	\end{subfigure}}
\caption{The comparison results of DSwMPC, DeSwMPC and CSwMPC in Example 1 in terms of the states trajectories.}
\label{Fig9}
\end{figure*}

\begin{figure*}[ht]
  \subfloat[subsystem 1]{
	\begin{subfigure}[c][1\width]{0.4\textwidth}
	   \centering
	   \includegraphics[width=1\textwidth]{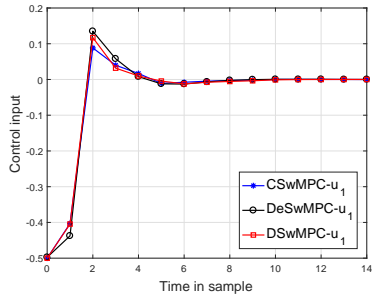}
	\end{subfigure}}
 \hfill 	
  \subfloat[subsystem 2]{
	\begin{subfigure}[c][1\width]{0.4\textwidth}
	   \centering
	   \includegraphics[width=1\textwidth]{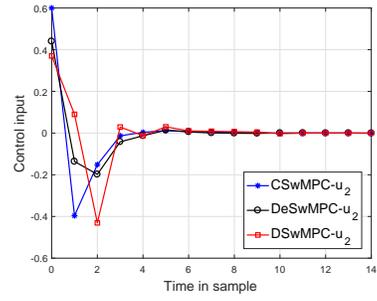}
	\end{subfigure}}
 \hfill	
  \subfloat[subsystem 3]{
	\begin{subfigure}[c][1\width]{0.4\textwidth}
	   \centering
	   \includegraphics[width=1\textwidth]{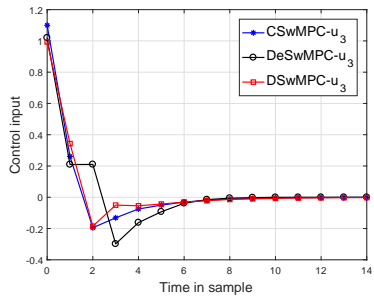}
	\end{subfigure}}
 \hfill	
  \subfloat[subsystem 4]{
	\begin{subfigure}[c][1\width]{0.4\textwidth}
	   \centering
	   \includegraphics[width=1\textwidth]{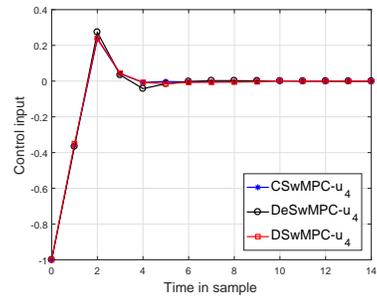}
	\end{subfigure}}
\caption{The comparison results of DSwMPC, DeSwMPC and CSwMPC in Example 2 in terms of the input trajectories. }
\label{Fig10}
\end{figure*}

\begin{figure*}[ht]
  \subfloat[subsystem 1]{
	\begin{subfigure}[c][1\width]{0.4\textwidth}
	   \centering
	   \includegraphics[width=1\textwidth]{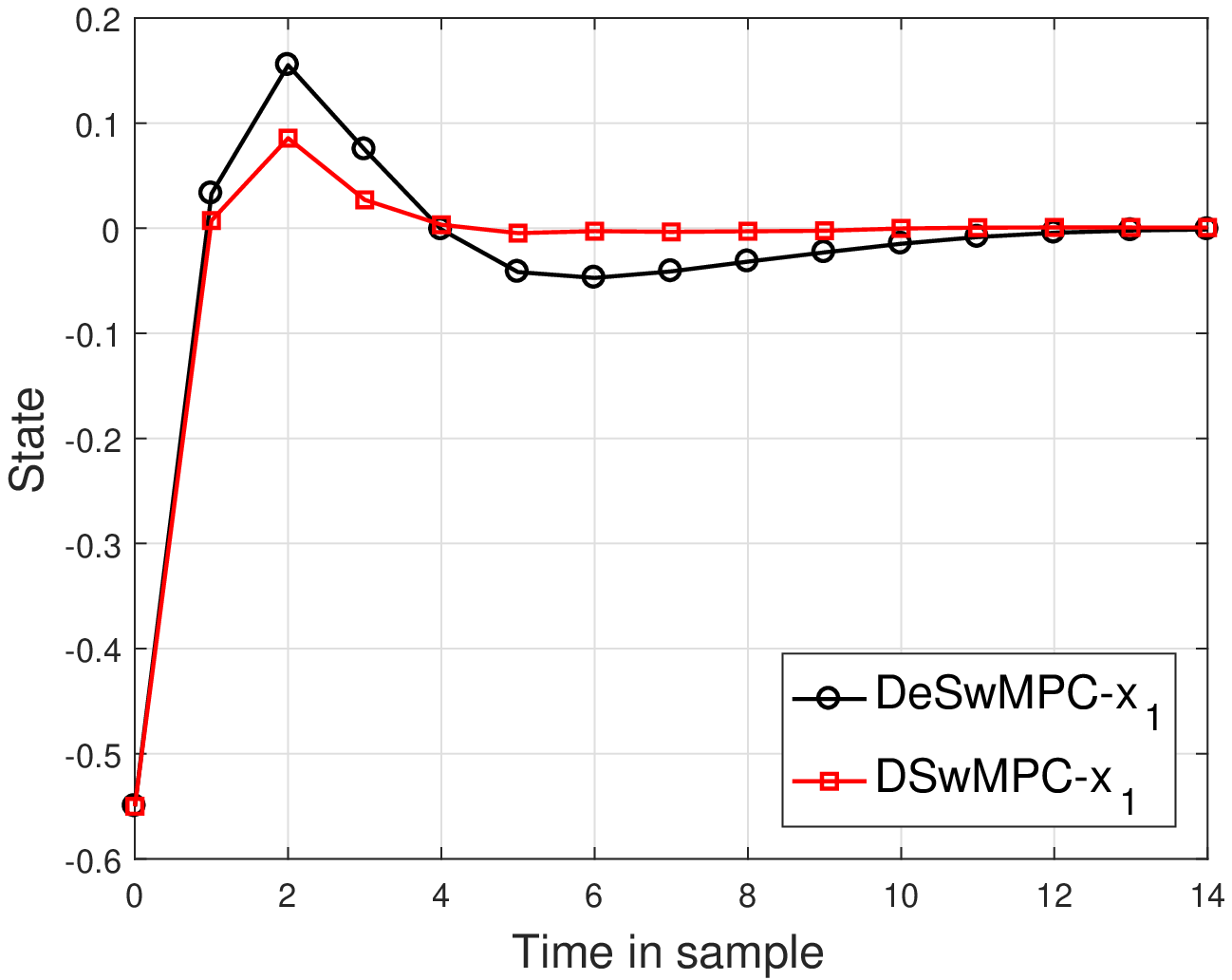}
	\end{subfigure}}
 \hfill 	
  \subfloat[subsystem 1]{
	\begin{subfigure}[c][1\width]{0.4\textwidth}
	   \centering
	   \includegraphics[width=1\textwidth]{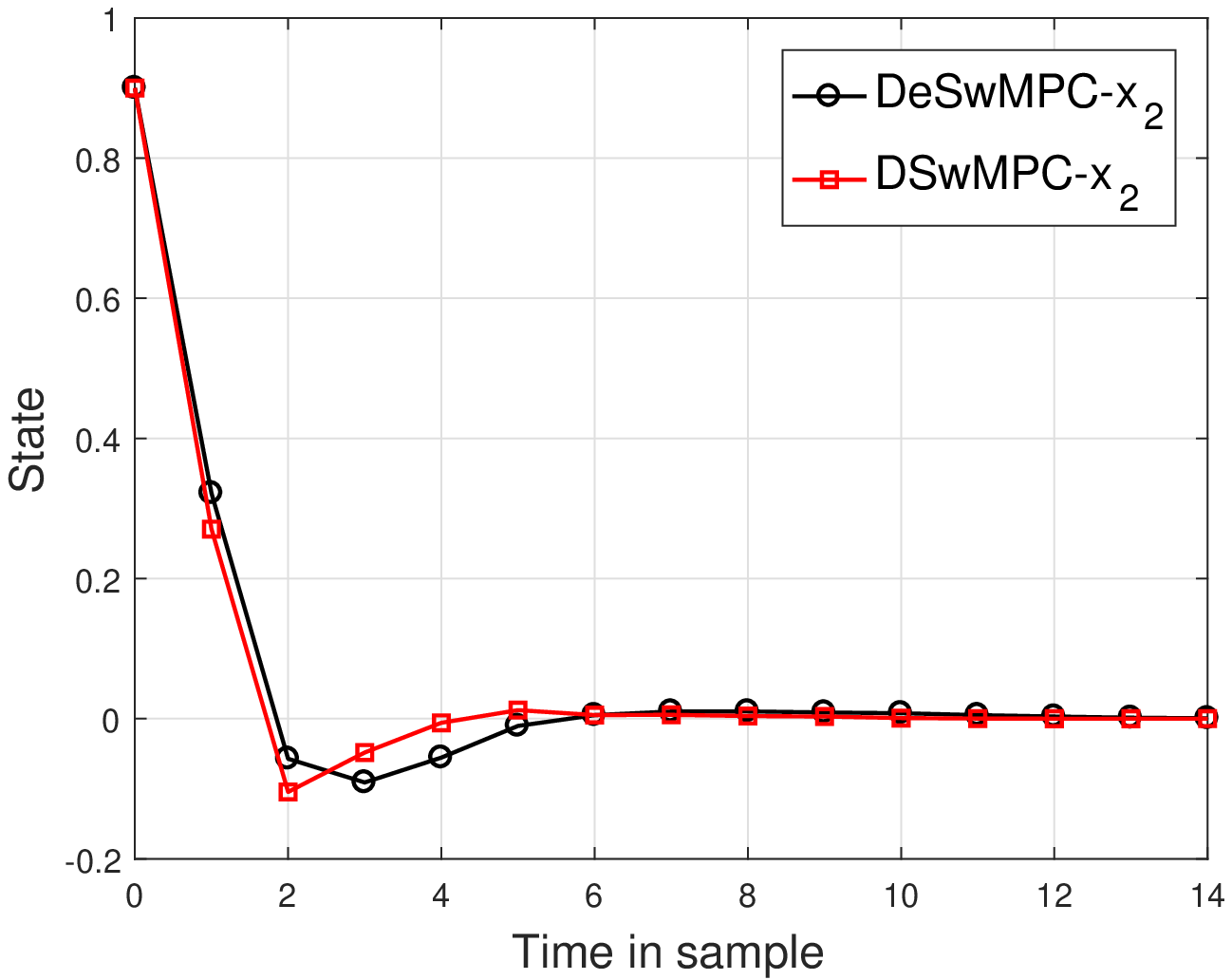}
	\end{subfigure}}
 \hfill	
  \subfloat[subsystem 2]{
	\begin{subfigure}[c][1\width]{0.4\textwidth}
	   \centering
	   \includegraphics[width=1\textwidth]{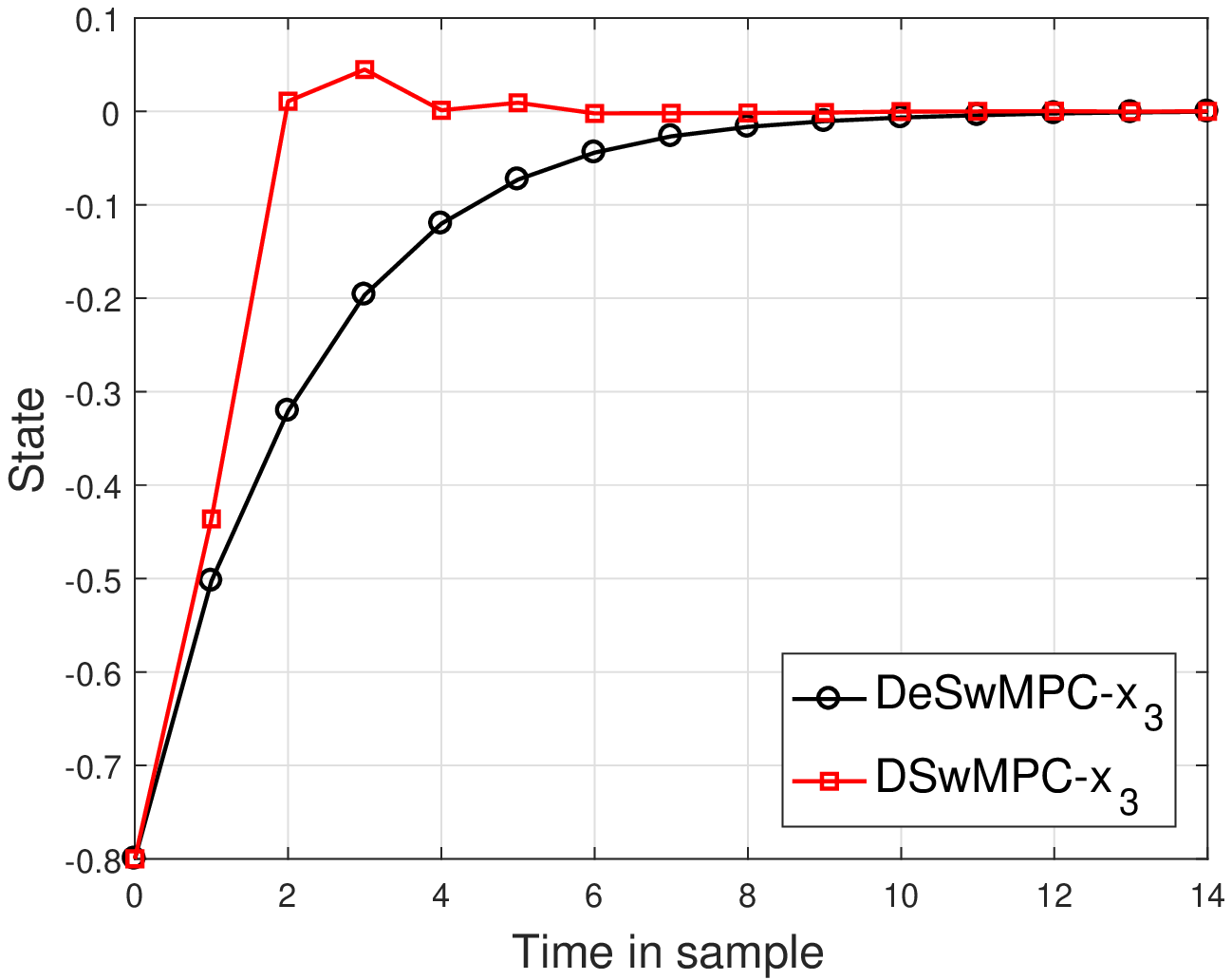}
	\end{subfigure}}
 \hfill	
  \subfloat[subsystem 2]{
	\begin{subfigure}[c][1\width]{0.4\textwidth}
	   \centering
	   \includegraphics[width=1\textwidth]{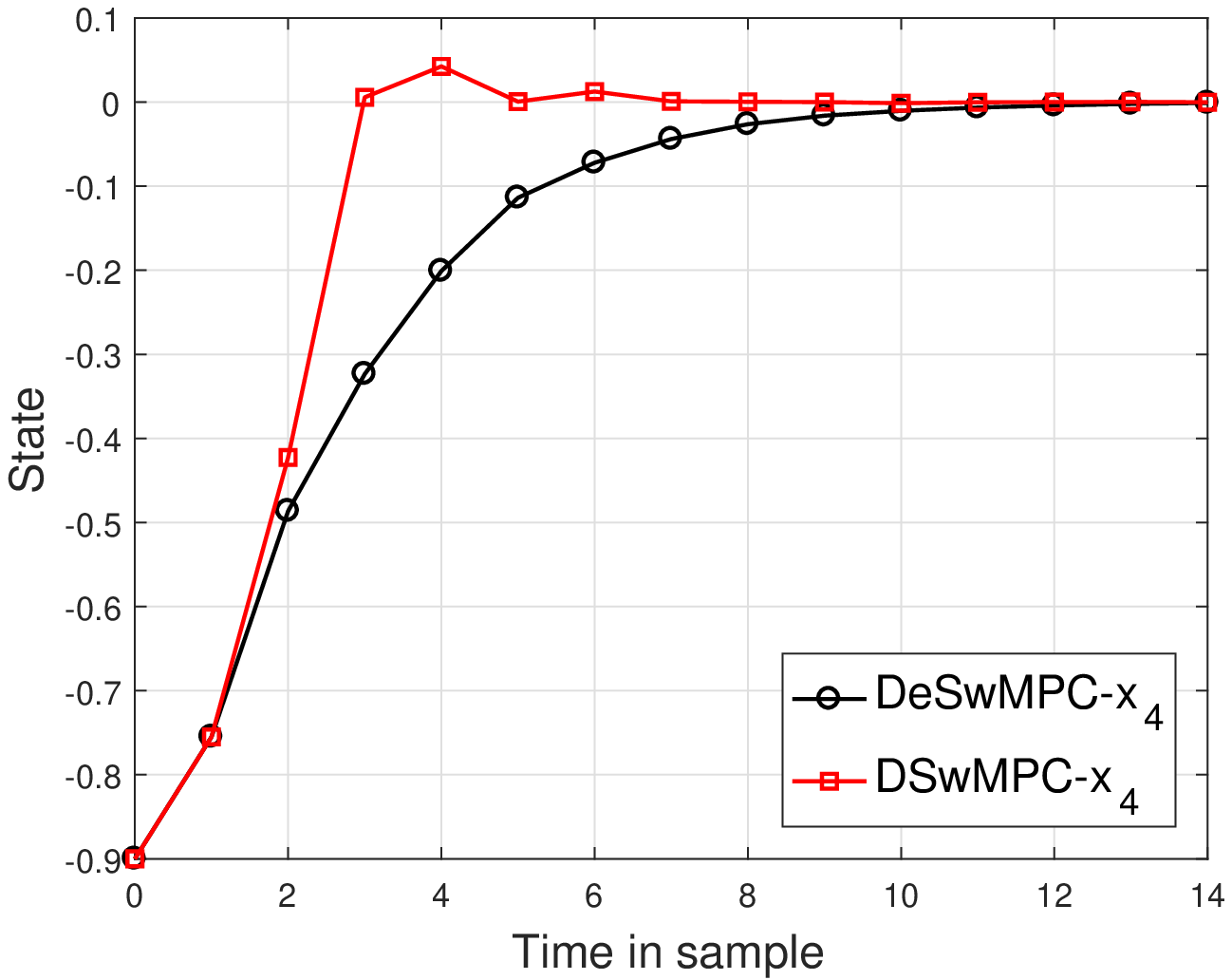}
	\end{subfigure}}
 \hfill	
  \subfloat[subsystem 3]{
	\begin{subfigure}[c][1\width]{0.4\textwidth}
	   \centering
	   \includegraphics[width=1\textwidth]{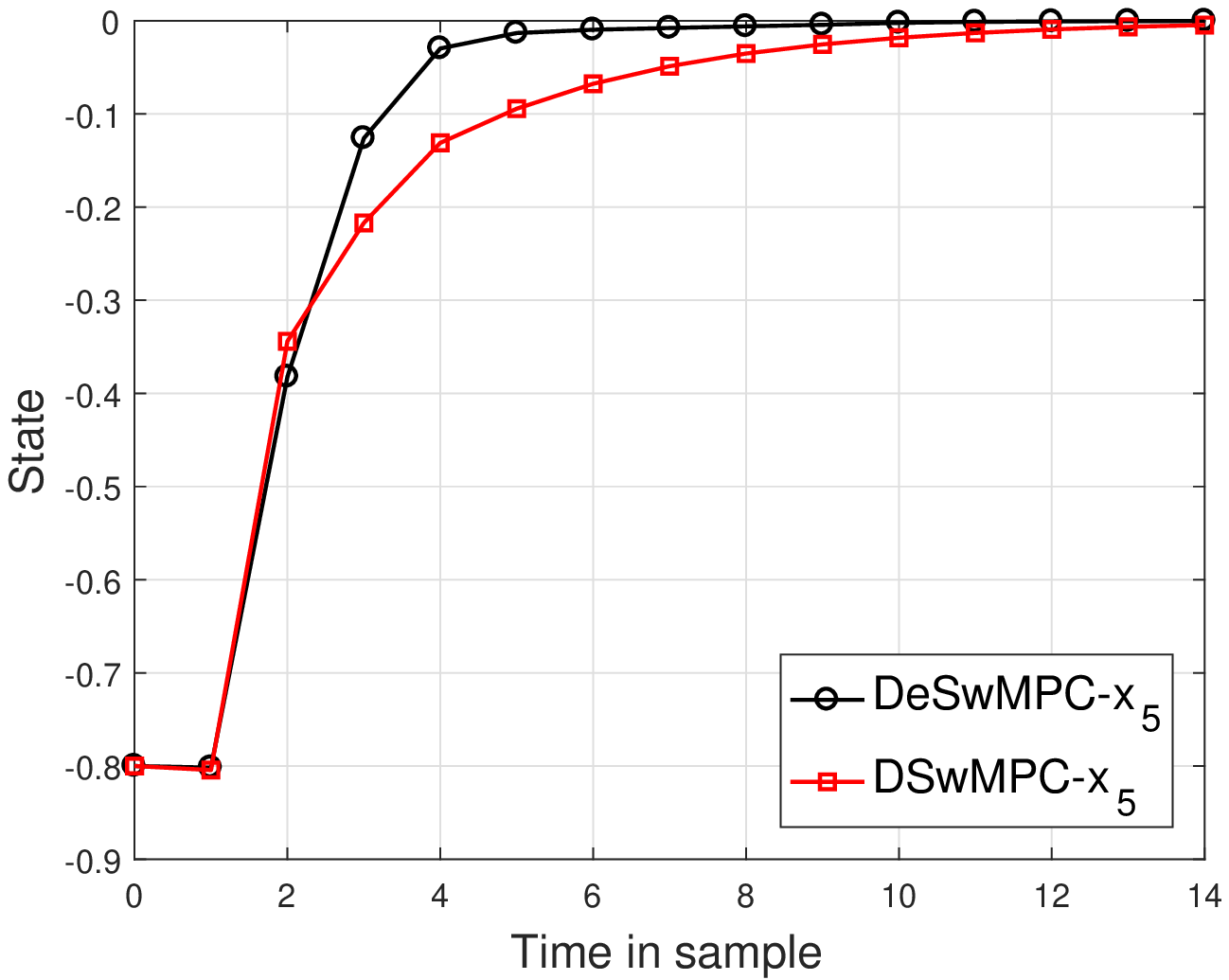}
	\end{subfigure}}
 \hfill	
  \subfloat[subsystem 3]{
	\begin{subfigure}[c][1\width]{0.4\textwidth}
	   \centering
	   \includegraphics[width=1\textwidth]{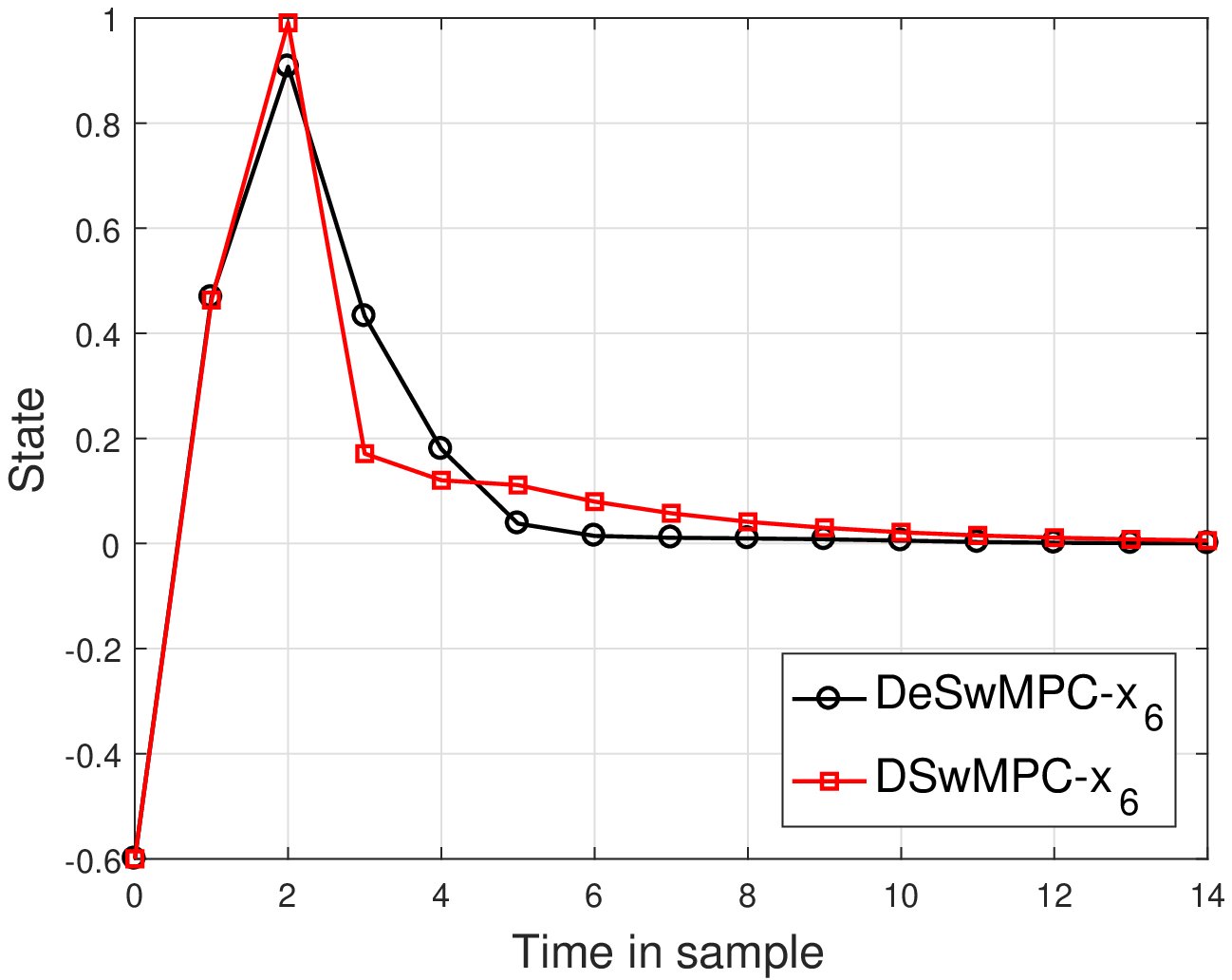}
	\end{subfigure}}
\end{figure*}
\begin{figure*}[ht]
\ContinuedFloat	
  \subfloat[subsystem 4]{
	\begin{subfigure}[c][1\width]{0.4\textwidth}
	   \centering
	   \includegraphics[width=1\textwidth]{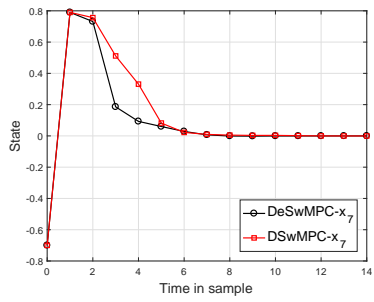}
	\end{subfigure}}
\hfill
  \subfloat[subsystem 4]{
	\begin{subfigure}[c][1\width]{0.4\textwidth}
	   \centering
	   \includegraphics[width=1\textwidth]{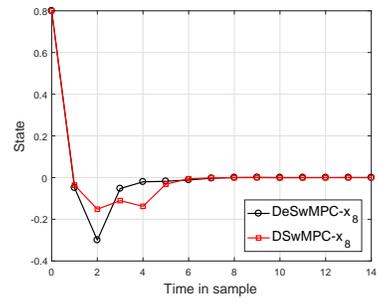}
	\end{subfigure}}
\caption{The comparison results of DSwMPC and DeSwMPC in Example 2 in terms of the states trajectories.}
\label{Fig10}
\end{figure*}

\begin{figure*}[ht]
  \subfloat[subsystem 1]{
	\begin{subfigure}[c][1\width]{0.4\textwidth}
	   \centering
	   \includegraphics[width=1\textwidth]{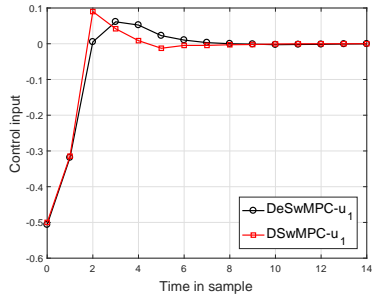}
	\end{subfigure}}
 \hfill 	
  \subfloat[subsystem 2]{
	\begin{subfigure}[c][1\width]{0.4\textwidth}
	   \centering
	   \includegraphics[width=1\textwidth]{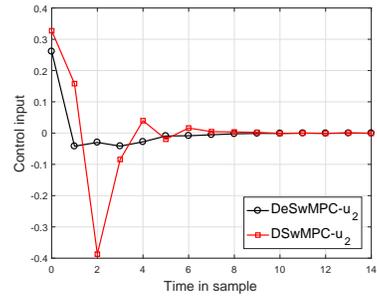}
	\end{subfigure}}
 \hfill	
  \subfloat[subsystem 3]{
	\begin{subfigure}[c][1\width]{0.4\textwidth}
	   \centering
	   \includegraphics[width=1\textwidth]{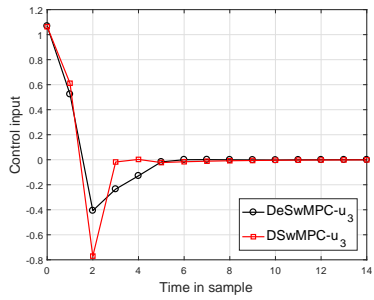}
	\end{subfigure}}
 \hfill	
  \subfloat[subsystem 4]{
	\begin{subfigure}[c][1\width]{0.4\textwidth}
	   \centering
	   \includegraphics[width=1\textwidth]{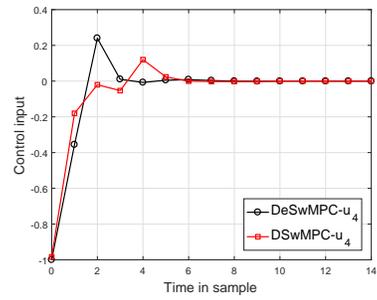}
	\end{subfigure}}
\caption{The comparison results of DSwMPC and DeSwMPC in Example 1 in terms of the input trajectories. }
\label{Fig12}
\end{figure*}

\begin{table*}
\caption{Sum state square errors of CSwMPC, DeSwMPC and DSwMPC on Examples 1-3.}
\label{Table3}
\begin{center}
\begin{small}
\begin{tabular}{c p{0.1cm} c  p{0.1cm} c  p{0.1cm} c}
  \hline
    Method & & {Example 1} & & {Example 2} & & {Example 3}  \\
             \hline
  CSwMPC   & & 8.1124 & & - & & -  \\
  DeSwMPC  & & 8.2829 & & 8.5480 & & -  \\
  DSwMPC   & & 8.3129 & & 8.4326 & & 8.8579  \\
  \hline
\end{tabular}
\end{small}
\end{center}
\end{table*}

\section{Conclutions} \label{Sect7:Conclutions}
A DSwMPC was presented in this study for handling dynamic interconnections among subsystems in a distributed switched large-scale system subjected to state/input constraints. In a distributed switched large-scale system, coupling among subsystems varies over time according to a switching signal. Such systems are described as the networked control systems with time-varying network topology in which a switching signal determines the neighborhood set of each subsystem over time. The proposed DSwMPC ensured the stability of the origin of the whole closed-loop system and as well constraints satisfaction. In the employed DMPC, by modeling the coupling terms of subsystems as the additive disturbance, their effect appeared in the dynamic equation, local states and input control constraints of the nominal subsystem. The local controllers exploited tube-based MPC to guarantee robustness than interconnections among subsystems. To consider the effect of switching signal which creates a time-varying network topology, the DSwMPC employed a robust tube-based SwMPC with the switch-RCI set as the target set robust to unknown mode switching as local regulators. Since the switch of network topology changes interactions among subsystems, it affects the terms of future state and input reference trajectories and additive disturbance in the dynamic of each subsystem, so the switching signal affects the dynamic equation and constraints of nominal subsystems.
Three typical examples were employed to illustrate the merits and effectiveness of DSwMPC. In all of them, it was supposed that the switching times are unknown in prior, but the next neighborhood sets were assumed to be known in prior in the first example and it was assumed to be unknown in the second and third ones. The simulation results demonstrated that the proposed DSwMPC satisfied the input and state constraints at all times. They also validated that the closed-loop system converges to the origin. It can be seen from the comparison results that,compared to the CSwMPC and the DeSwMPC, the DSwMPC proposed in this article has a flexible structure to be applied on distributed switched systems in which the system topology changes under various scenarios.

\bibliographystyle{elsarticle-num}
\bibliography{Paper}

\end{document}